\documentclass[fleqn,usenatbib]{mnras}

\usepackage{newtxtext,newtxmath}

\usepackage[T1]{fontenc}
\usepackage{ae,aecompl}
\usepackage{pdflscape}
\usepackage{soul}
\usepackage[normalem]{ulem}
\usepackage[usenames,dvipsnames]{color}

\usepackage{graphicx}	
\usepackage{amsmath}	
\usepackage{amssymb}	

\newcommand{\mvec}[1]{\boldsymbol{#1}}

\newcommand{\Msun}{\mathrm{M}_\odot}
\newcommand{\Msunh}{\mathrm{M}_\odot h^{-1}}
\newcommand{\pkg}[1]{\texttt{#1}}
\newcommand{\single}[1]{\overline{#1}}

\newcommand{\model}[1]{\texttt{#1}}

\newcommand{\smag}{$C_\mathrm{s}$ }

\newcommand{\xmark}{$\times$}
\newcommand{\tmix}{\tau_\mathrm{mix}}
\newcommand{\Zsun}{\mathrm{Z}_\mathrm{\odot}}

\newcommand{\kpch}{\mathrm{kpc}\, h^{-1}}
\newcommand{\kpc}{\mathrm{kpc}}
\newcommand{\Mpc}{\mathrm{Mpc}}
\newcommand{\cMpch}{\mathrm{cMpc}\,h^{-1}}

\title[Sub-grid turbulence]{Mixing matters}

\author[D. Rennehan]{
Douglas Rennehan$^{1}$\thanks{E-mail: douglas.rennehan@gmail.com (DR)}
\\
$^{1}$Department of Physics \& Astronomy, University of Victoria, BC V8P 5C2, Canada
}

\date{Accepted XXX. Received YYY; in original form ZZZ}

\pubyear{2021}

\begin{document}
\label{firstpage}
\pagerange{\pageref{firstpage}--\pageref{lastpage}}
\maketitle

\begin{abstract}
All hydrodynamical simulations of turbulent astrophysical phenomena require sub-grid scale models to properly treat energy dissipation and metal mixing.  We present the first implementation and application of an anisotropic eddy viscosity and metal mixing model in Lagrangian astrophysical simulations, including a dynamic procedure for the model parameter.  We compare these two models directly to the common Smagorinsky and dynamic variant.  Using the mesh-free finite mass method as an example, we show that the anisotropic model is best able to reproduce the proper Kolmogorov inertial range scaling in homogeneous, isotropic turbulence.  Additionally, we provide a method to calibrate the metal mixing rate that ensures numerical convergence.  In our first application to cosmological simulations, we find that all models strongly impact the early evolution of galaxies leading to differences in enrichment and thermodynamic histories.  The anisotropic model has the strongest impact, with little difference between the dynamic and the constant-coefficient variant.  We also find that the metal distribution functions in the circumgalactic gas are significantly tighter at all redshifts, with the anisotropic model providing the tightest distributions.  This is contrary to a recent study that found metal mixing to be relatively unimportant on cosmological scales.  In all of our experiments the constant-coefficient Smagorinsky and anisotropic models rivaled their dynamic counterparts, suggesting that the computationally inexpensive constant-coefficient models are viable alternatives in cosmological contexts.
\end{abstract}




\section{Introduction}

Galaxies form and evolve in tempestuous gaseous environments where hydrodynamics, radiative cooling, and gravity synergize to produce rich emergent phenomena on a myriad of spatial scales.  The immense dynamic range of scales involved and their interconnectedness prove to be limiting factors in advancing our understanding of the complete picture of galaxy evolution (see \citealt{Naab2016} for an excellent review).  At the forefront of the issue is hydrodynamical turbulence as it is a multi-scale, non-linear phenomenon that occurs in almost all galactic environments --- directly impacting our theoretical understanding of galaxy evolution.  

While the importance of turbulence in the interstellar medium of galaxies has been long recognized (see \citealt{Elmegreen2004} and \citealt{Scalo2004} for reviews), only recently has the role of turbulence in halo gas come under careful consideration.  Indeed, both the circumgalactic medium (CGM) of $L^*$ galaxies and the intracluster medium (ICM) of groups and clusters of galaxies show signs of turbulence playing an important role in their evolution (see, for example, \citealt{Prasad2018} and \citealt{Wang2020}).  Observationally, there is evidence of complex kinematic structure in the CGM of $L^*$-galaxies that emerged through the revolutionary \textit{Cosmic Origins Spectrograph} halo survey on the Hubble space telescope (\textit{COS}-halos; \citealt{Tumlinson2013, Tumlinson2017}).   For instance, \cite{Werk2016} found that turbulent velocities of $50$-$75$ $\mathrm{km}\,\mathrm{s}^{-1}$ explain the broadening of absorption lines in the CGM that is not otherwise explainable that has subsequently been confirmed using numerical studies \citep{Buie2020}.   Moving up in mass scale, the ICM also shows evidence of turbulence through indirect observational methods such as X-ray surface brightness and Sunyaev-Zeldovich fluctuations \citep{Zhuravleva2014, Pinto2015, Zhuravleva2015, Khatri2016, Zhuravleva2018}.

There are two main drivers of turbulence in galactic environments: (a) global outflows that emerge from star formation processes and supermassive black holes (SMBHs) within galaxies \citep{Prasad2015, Prasad2018, Yang2016, Bourne2017, Fielding2017, Fielding2018, Sokolowska2018, Li2020b} and (b) shearing motion driven by gas in-fall during structure formation \citep{Dekel2009, Vazza2010, Vazza2012, Vazza2017, Wittor2017, Bennett2020}, mergers \citep{Zuhone2013}, and ram-pressure stripping of galaxies moving through the ICM \citep{Ruggiero2016, Simons2020}.  In both the CGM and ICM, turbulence could provide additional pressure support \citep{Poole2006, Vazza2018, Lochhaas2020} that prevents the gas from collapsing and rapidly converting into stars as well as a physical mechanism to transport energy and metals throughout gas directly -- impacting the cooling profile, star formation cycle, and metal distribution functions \citep{Shen2010, Shen2012, Shen2013, Brook2014, Sokolowska2018, Escala2017, Tremmel2018,  Rennehan2019, Hafen2019, Hafen2020}.  Therefore, understanding the nature of turbulence is imperative to understand the complete picture of galaxy evolution.

While there are many successful cosmological simulations that use a variety of sub-grid assumptions to broadly reproduce galaxy populations (e.g. \citealt{Guedes2011, Hopkins2014, Vogelsberger2014, Schaye2014, Genel2014, Dave2017a, Dave2019, Tremmel2018, Pillepich2018, Huang2020a}), one aspect that is often overlooked is the numerical modelling of sub-grid turbulence.  The crux of the problem is in the fact that in hydrodynamical simulations, the physical dissipation scale $\eta$ is almost always much smaller than the resolution scale, $\eta \ll h$ \citep{Pope2000}.  For that reason, the kinetic energy that is flowing in the turbulent cascade reaches some scale $H \gtrsim h$ where it may no longer progress.  If the numerical viscosity of the hydrodynamical method is not sufficiently strong to thermalise the kinetic energy, there will be a build-up of kinetic energy at that scale $H \gtrsim h \gg \eta$ \citep{Garnier2009}.  The kinetic energy build-up is a completely unphysical representation of turbulence and not only impacts the energetics, but the large scale flow properties such as the redistribution of metals.  

Although not explicitly stated, many cosmological hydrodynamical simulation studies implicitly assume that \textit{numerical dissipation} is sufficient to mimic sub-grid turbulence, rather than modelling  sub-grid turbulence with additional terms in the hydrodynamical equations of motion.  However, numerical dissipation is not sufficient to reduce the kinetic energy build-up and may not represent turbulent flow statistics in all cases \citep{Sagaut2006}.  Indeed, \cite{Lecoanet2016} showed that numerical noise at the resolution scale seeds instability in the Kelvin-Helmholtz experiment that causes the long-term evolution to be unconverged as the small-scale instabilities grow.  When they introduced explicit sub-grid diffusion to their simulations, the results of the Kelvin-Helmholtz experiment converged -- showing that more small-scale structure (i.e. resolution-scale noise) is not necessarily better, and that explicit sub-grid diffusion can cause \textit{less} large-scale mixing.

In simulations that use Eulerian (i.e. grid-based) hydrodynamics, there has been extensive effort to developing models for sub-grid turbulence (see \citealt{Sagaut2006, Garnier2009} and \citealt{Schmidt2015} for extensive reference lists).  For astrophysically-relevant work, we refer the reader to \cite{Scannapieco2008}, \cite{Pan2013}, \cite{Federrath2013a}, \cite{Schmidt2013}, and \cite{Semenov2015} (for a review, see \citealt{Schmidt2015}).  In this paper, we focus on the Lagrangian hydrodynamical methods.

In Lagrangian hydrodynamics, the fluid equations are approximated via fluid elements that move with the flow.  There are three main approaches that are most common in cosmological simulations: smoothed particle hydrodynamics (SPH) \citep{Gingold1977, Lucy1977, Hernquist1989, Hopkins2013}, the moving-mesh method (MM) \citep{Springel2010}, and mesh-free methods (MF) \citep{Lanson2008a, Lanson2008b, Gaburov2011, Hopkins2015a}.  Each of these methods track fluid elements using different discretisation techniques that lead to different levels of numerical dissipation.  In SPH, there is no inherent numerical dissipation and it, counter-intuitively, produces a deficit of kinetic energy near the resolution scale rather than a build-up \citep{Bauer2012, Price2012}.  The MM and MF methods use Riemann solvers to approximate the fluid equations of motion between neighbouring fluid elements.  Riemann solvers are generally diffusive due to their approximate nature, and the build-up of kinetic energy is present in both methods on scales up to $\sim10$ times the resolution scale \citep{Bauer2012, Hopkins2015a}.  

A solution to kinetic energy build-up is to model the action of turbulent eddies as a viscous process that diffuses momentum (and metals) and dissipates kinetic energy via a diffusion equation and source term in the energy equation, respectively.  Usually the assumption is that the viscosity, or diffusivity, depends on velocity fluctuations $v_\mathrm{eddy}$ near the resolution scale $h$.  The resulting diffusivity is $D \propto hv_\mathrm{eddy}$, where $v_\mathrm{eddy}$ may be a characteristic velocity, or velocity difference, within the neighbourhood of a fluid element \citep{Wadsley2008, Greif2009a}.  

One particularly important choice of $v_\mathrm{eddy}$ is the Smagorinsky model \citep{SMAGORINSKY1963}, which assumes that velocity shear fluctuations drive dissipation and mixing through $v_\mathrm{eddy} \sim h |S^{*}|$, where $|S^{*}|$ is the magnitude of the trace-free shear tensor.  The Smagorinsky model has been successfully used to treat metal and thermal energy mixing in SPH \citep{Wadsley2008, Shen2010, Shen2012, Brook2014, Williamson2016a, Tremmel2017a, Wadsley2017, Su2016} and has been extended to other Lagrangian hydrodynamical methods, such as the MFM method \citep{Colbrook2017, Escala2017,  Rennehan2019, Hafen2019, Hafen2020}.  While the Smagorinsky model improves mixing in Lagrangian simulations, it is important to consider two assumptions in the model: (a) that shear fluctuations (i.e., changes in $|S^*|$) always represent turbulence and (b) that the diffusive process acts isotropically through the magnitude of the trace-free (i.e. ignoring compression) shear tensor over the scale $h$.  Are either of these assumptions reasonable?

To address point (a),  \cite{Piomelli1994} proposed a method of dynamically calculating the model coefficient at simulation time.  That method was employed in \cite{Rennehan2019} for the first time in Lagrangian hydrodynamics and they found that the model significantly reduced over-diffusion in non-turbulent shear flows, such as in rotating galactic disks and the Kelvin-Helmholtz instability.

The second point (b) concerns the isotropy in $D$ and the discounting of compression in $|S^*|$.  Recently \cite{Hu2020} showed that a better representation of sub-grid scale turbulence is obtained by treating the full velocity tensor $\mvec{\nabla}\otimes\mvec{u}$ for the diffusivity $\mvec{D}$, now a tensor.  The model is the \textit{gradient} model \citep{Clark1979} and is a major change since diffusion now depends on the directionality encoded in $\mvec{\nabla}\otimes\mvec{u}$ rather than equally in every spatial direction.  The trace of $\mvec{\nabla}\otimes\mvec{u}$ is automatically included and, therefore, compression is automatically handled\footnote{The trace of $\mvec{\nabla}\otimes\mvec{u}$ is $\mvec{\nabla}\cdot\mvec{u}$, the divergence of the flow.} -- an important point for highly-compressible turbulence in cosmological flows.  However, \cite{Hu2020} post-processed their driven turbulence simulations to check if the model \textit{would have} improved the results at simulation time.  Our goal is to implement the model for the mesh-free finite mass method, and determine its feasibility \textit{at simulation time} in combination with the dynamic procedure from \cite{Rennehan2019}.

We introduce an implementation of the gradient model for Lagrangian astrophysical simulations and additionally provide methods for computing the model coefficient at simulation time.  In Section~\ref{sec:methods} we provide a derivation of the model, as well as a derivation of the dynamic procedure in Section~\ref{sec:methods_dynamic_model} that allows calculation of the model coefficient at simulation time.  Section~\ref{sec:driven_turbulence} describes driven turbulence validation tests of the gradient model, included at run time, for both eddy viscosity and metal mixing.  As a first application, we describe the qualitative impact of the eddy viscosity and metal mixing model on cosmological gas phases in Section~\ref{sec:cosmological}.  We present our conclusions and recommendations in Section~\ref{sec:conclusions}.

\section{The gradient model}
\label{sec:methods}

In finite-mass Lagrangian hydrodynamics, such as the mesh-free finite mass (MFM) method \citep{Lanson2008a, Lanson2008b, Gaburov2011, Hopkins2015a}, the build-up of kinetic energy at the resolution scale demands an additional dissipation mechanism.  Additionally, metals follow the fluid mass elements throughout the simulation volume and, therefore, the exchange of metals between fluid elements due to sub-grid scale turbulent motion does not occur.  The crux of the issue is that discretisation of the fluid field leads to damping out of the high-frequency turbulent fluctuations that should continue down to the \textit{physical} dissipation scale.  It is useful to think of the damping action as a high-pass filter acting on the fluid equations of motion.  By applying a general filter to the momentum conservation equation, it is possible to derive the correct level of mixing that should occur between fluid elements due to unresolved turbulence. 

In general, the filtering action over a scalar field $f_i(\mvec{r})$ can be represented as,

\begin{equation}
\label{eq:filteroperation}
\overline{f}_{i}(\mvec{r}) = \int_{D} f_i(\mvec{r}') G(|\mvec{r} - \mvec{r}'|, h) d\mvec{r}',
\end{equation}

\noindent where $h$ is the smoothing scale over the domain $D$.  We apply this to the momentum equation to determine the correction terms due to unresolved turbulence,

\begin{equation}
\label{eq:momentumtransport}
\frac{\partial (\rho \mvec{u})}{\partial t} + \nabla\cdot(\rho \mvec{u} \otimes \mvec{u} + P\mvec{I}) = 0,
\end{equation}

\noindent where $\rho$ is the gas density, $P$ is the pressure, and $\mvec{u}$ is the velocity vector.  If we apply equation~(\ref{eq:filteroperation}) to equation~(\ref{eq:momentumtransport}) it follows that, assuming the filtering operation and derivatives commute,

\begin{equation}
\label{eq:filtermomentum}
\frac{\partial (\overline{\rho \mvec{u}})}{\partial t} + \nabla\cdot(\overline{\rho \mvec{u} \otimes \mvec{u}} + \overline{P}\mvec{I}) = 0.
\end{equation}

\noindent For simplicity, we switch to density-weighted variables such that $\widetilde{\mvec{u}} \equiv \overline{\rho \mvec{u}} / \overline{\rho}$ and

\begin{equation}
\label{eq:tildemomentum}
\frac{\partial (\overline{\rho} \tilde{\mvec{u}})}{\partial t} + \nabla\cdot(\overline{\rho} \widetilde{\mvec{u}\otimes\mvec{u}} + \overline{P}\mvec{I}) = 0.
\end{equation}

\noindent The term $\widetilde{\mvec{u}\otimes\mvec{u}}$ is unknown at simulation time because it relies on information below the resolution scale.  To put the equation in a more manageable form, we add $\nabla\cdot(\overline{\rho}[\widetilde{\mvec{u}}\otimes\widetilde{\mvec{u}}-\widetilde{\mvec{u}\otimes\mvec{u}}])$ and rearrange,

\begin{equation}
\label{eq:filtermomentum3}
\frac{\partial (\overline{\rho} \widetilde{\mvec{u}})}{\partial t} + \nabla\cdot(\overline{\rho} \widetilde{\mvec{u}}\otimes\widetilde{\mvec{u}} + \overline{P}\mvec{I}) = \nabla\cdot(\overline{\rho} [\widetilde{\mvec{u}}\otimes\widetilde{\mvec{u}} - \widetilde{\mvec{u}\otimes\mvec{u}}]).
\end{equation}

\noindent Therefore, we define the sub-grid scalar flux $\mvec{F}$,

\begin{equation}
\label{eq:scalarflux}
\mvec{F} \equiv \overline{\rho} (\widetilde{\mvec{u}\otimes\mvec{u}} - \widetilde{\mvec{u}}\otimes\widetilde{\mvec{u}}),
\end{equation}

\noindent and retrieve a new equation,

\begin{equation}
\label{eq:momentumtransportfinal}
\frac{\partial (\overline{\rho} \widetilde{\mvec{u}})}{\partial t} + \nabla\cdot(\overline{\rho} \widetilde{\mvec{u}}\otimes\widetilde{\mvec{u}} + \overline{P}\mvec{I}) = -\nabla\cdot\mvec{F}.
\end{equation}

\noindent The sub-grid scale momentum flux $\mvec{F}$ is unknown at simulation time and must be modelled yet is widely ignored in cosmological simulation studies which usually focus only on the thermal energy and metal fluxes via the Smagorinsky model.

There are a myriad of models in the literature for $\mvec{F}$ (see \citealt{Garnier2009} for extensive lists)  but here we take the direct approach of using a Taylor series approximation following \cite{Clark1979} and \cite{Hu2020}.  We expand\footnote{See Appendix \ref{app:filterapprox} for a full derivation.} $\widetilde{\mvec{u}\otimes\mvec{u}}$ via a Taylor expansion as

\begin{equation}
\widetilde{\mvec{u}\otimes\mvec{u}} \approx \widetilde{\mvec{u}}\otimes\widetilde{\mvec{u}} + \epsilon \nabla^2(\widetilde{\mvec{u}}\otimes\widetilde{\mvec{u}})
\end{equation}

\noindent and, therefore, the flux becomes

\begin{equation}
\label{eq:fluxdef}
\mvec{F} = \overline{\rho} (\widetilde{\mvec{u}\otimes\mvec{u}} - \widetilde{\mvec{u}}\otimes\widetilde{\mvec{u}}) \approx \overline{\rho} \epsilon \nabla^2(\widetilde{\mvec{u}}\otimes\widetilde{\mvec{u}}),
\end{equation}

\noindent where $\epsilon$ is a constant that depends on the kernel scale $h$ as $\epsilon \propto h^2$ \citep{Monaghan1989, Monaghan2002, Monaghan2011}.  Expanding the term on the right hand side of equation~(\ref{eq:fluxdef}) and keeping only the first derivative terms we find

\begin{equation}
\label{eq:realflux}
\mvec{F} = 2\overline{\rho}Ch^2 (\nabla\otimes\widetilde{\mvec{u}})(\nabla\otimes\widetilde{\mvec{u}})^{\mathrm{T}},
\end{equation}

\noindent where $C$ is our model parameter.

A similar result emerges when considering the mass flux of metals in a fluid,

\begin{equation}
\label{eq:metaltransport}
\frac{\partial (\rho Z)}{\partial t} + \mvec{\nabla}\cdot(\rho \mvec{u} Z) = 0.
\end{equation}

\noindent Applying the same approach as before gives an equation for the sub-grid flux of metals, 

\begin{equation}
\label{eq:metalflux}
\mvec{F} = 2\overline{\rho}C_\mathrm{Z}h^2 (\nabla\otimes\widetilde{\mvec{u}})\cdot\mathbf{\nabla}Z
\end{equation}

The method for solving equation~(\ref{eq:momentumtransportfinal}) is detailed in \cite{Hopkins2016b}, and we point the reader to that work for further information.  From their equation (2),

\begin{equation}
\label{eq:hopkins2}
\mvec{F} = \mvec{K}\cdot(\nabla\otimes\mvec{q}),
\end{equation}

\noindent where $\mvec{K}$ is the tensor describing the diffusive strength, and $\mvec{q}$ is the fluid field property.  Therefore, we identify,

\begin{equation}
\mvec{K} \equiv 2\overline{\rho} Ch^2 (\nabla\otimes\widetilde{\mvec{u}}),
\end{equation}
\begin{equation}
\mvec{q} \equiv \widetilde{\mvec{u}}.
\end{equation}

\noindent It is important to note that following Section 3.0.6 of \cite{Hopkins2016b}, we also include the dissipation term corresponding to $\mvec{\nabla}\cdot\mvec{F}$ in the energy flux to ensure energy conservation.

The model in equation~(\ref{eq:realflux}) is known to lead to numerical instability due to particles attracting rather than repelling \citep{Nomura1998, Balarac2013}, similar to the well-studied tensile instability in smoothed particle magnetohydrodynamics \citep{Phillips1985, Morris1996, Monaghan2000, Price2012a}.  \cite{Balarac2013} specifically showed for the anisotropic eddy viscosity model that ignoring the positive eigenvalues of $\mvec{S}$ ensures the model is always well behaved and that the action of the model is to repel particles rather than attract. Therefore, we follow \cite{Balarac2013} and only keep the contribution due to the negative eigenvalues of the shear tensor.  However, we must first decompose $\nabla\otimes\widetilde{\mvec{u}}$ into the symmetric and anti-symmetric parts,

\begin{equation}
\label{eq:veldecomp}
\nabla\otimes\widetilde{\mvec{u}} = \frac{1}{2}(\nabla\otimes\widetilde{\mvec{u}} + [\nabla\otimes\widetilde{\mvec{u}}]^\mathrm{T}) + \frac{1}{2}(\nabla\otimes\widetilde{\mvec{u}} - [\nabla\otimes\widetilde{\mvec{u}}]^\mathrm{T}) \equiv \mvec{S} + \mvec{\Omega}.
\end{equation}

\noindent We further decompose $\mvec{S}$ into the contribution from the positive eigenvalues and negative eigenvalues as $\mvec{S} \equiv \mvec{S}_\oplus + \mvec{S}_\ominus$. It then follows that

\begin{equation}
\label{eq:shear}
\mvec{S}_\ominus \equiv \sum_{k=1}^3 \mathrm{min}(0, \lambda^{(k)})\mvec{e}^{(k)}\otimes\mvec{e}^{(k)},
\end{equation}

\noindent where $\lambda^{(k)}$ is the $k$th eigenvalue and $\mvec{e}^{(k)}$ is the corresponding eigenvector.  Therefore, the new diffusion coefficient tensor is

\begin{equation}
\label{eq:finaldiffcoeff}
\mvec{K} = 2\overline{\rho}Ch^2\mvec{S}_\ominus,
\end{equation}

\noindent where we have reduced the shear contribution into $\mvec{S}_\ominus$.  Removing the negative eigenvalues appears to not be necessary for the metal mixing case as we found no cases of numerical instability in all of our hydrodynamical tests and cosmological simulations.  However, we find that it is absolutely necessary in cosmological simulations for the momentum flux.

It is important to note that our choice of $h$ differs from that in \cite{Hopkins2018}.  We choose $h$ as the kernel radius of compact support as this is the maximum interaction distance for the flux in the MFM method.  Physically, it is the maximum distance over which sub-grid eddies transport their fluid properties.  We find much better results in our tests in Section~\ref{sec:driven_turbulence} using the maximum interaction distance.  However, we do note that the radius of compact support is different for each kernel and may not be the most accurate length-scale.  Given that it is normally twice the smoothing scale for the kernel \citep{Dehnen2012}, it provides the best compromise.  Any kernel that is used having a compact support radius larger than twice the smoothing scale should be further investigated before using our definition of $h$.  Our value is approximately twice that found in the \pkg{FIRE} studies \citep{Su2016, Escala2017, Hafen2019, Hafen2020}, but produces $4$ times as much dissipation and metal mixing since the dependence is squared in equation~(\ref{eq:finaldiffcoeff}).  Our choice of $h$ is the same as \cite{Wadsley2017} who use the radius of compact support for turbulent mixing in the \pkg{GASOLINE-2} code.

\subsection{The dynamic gradient model}
\label{sec:methods_dynamic_model}

We apply the same procedure in \cite{Balarac2013} combined with the density-weighted filtering procedure in \cite{Rennehan2019}.  Following their notation, we replace $\tilde{f}$ with $\overline{f}$ since fluid properties are inherently density-weighted in the mesh-free finite mass method.  Although we focus on the velocity fluctuations in the following procedure, we note that it applies equally to the metal field.  

The resolved fluctuations in the flow are,

\begin{equation}
\label{eq:resolvedflux}
\mvec{\mathcal{L}} = \widehat{\mvec{\overline{u}}\otimes \overline{\mvec{u}}} - \widehat{\mvec{\overline{u}}} \otimes\widehat{\overline{\mvec{u}}},
\end{equation}

\noindent where $\widehat{\overline{\mvec{u}}}$ represents the velocity vector $\mvec{u}$ filtered once on the resolution scale (to produce $\overline{\mvec{u}}$), then filtered again on twice the resolution scale ($\widehat{h} \equiv \widehat{\overline{h}} \sim 2\overline{h}$, see Section 2.4 of \citealt{Rennehan2019}).  Explicitly, we represent the filtering operation on any scalar quantity $f$ from equation~(\ref{eq:filteroperation}) as a sum \citep{Monaghan1989, Monaghan2005, Monaghan2011, Rennehan2019},

\begin{equation}
\label{eq:smoothdiscrete}
\overline{f}_a = f_a + \epsilon \sum_b \frac{m_b}{\langle\rho_{ab}\rangle_{\single{h}}} (f_b - f_a) W(|\mvec{x_a} - \mvec{x_b}|, \overline{h}_{ab}),
\end{equation}

\noindent where $ f_a $ is any scalar quantity at particle $ a $, $ \overline{h}_{ab} $ is the mean smoothing lengths between the two particles, $\langle\rho_{ab}\rangle_{\single{h}}$ is the mean of the densities of the two particles, and the sum is taken over $ b $ nearest neighbours.  We take the smoothing factor $\epsilon = 0.8$ following \cite{Rennehan2019}.  To obtain double-filtered quantities, we apply equation~(\ref{eq:smoothdiscrete}) to the singly-filtered quantities,

\begin{equation}
\label{eq:smoothdiscrete2}
\widehat{\overline{f}}_a = \overline{f}_a + \epsilon \sum_b \frac{m_b}{\langle\rho_{ab}\rangle_{\widehat{h}}} (\overline{f}_b - \overline{f}_a) W(|\mvec{x_a} - \mvec{x_b}|, \widehat{h}_{ab}).
\end{equation}

If we use only resolved quantities (i.e. doubly-filtered) then the gradient model in equation~(\ref{eq:realflux}) should reproduce $\mvec{\mathcal{L}}$,

\begin{equation}
\mvec{\mathcal{L}} = \widehat{\mvec{\overline{u}}\otimes \overline{\mvec{u}}} - \widehat{\mvec{\overline{u}}}\otimes\widehat{\overline{\mvec{u}}} = 2 C \widehat{\overline{h}}^2 \widehat{\overline{\mvec{S}}}_\ominus (\mvec{\nabla}\otimes \widehat{\overline{\mvec{u}}})^{\mathrm{T}}.
\end{equation}

\noindent The above equation results in a solution for the one unknown parameter $C$,

\begin{equation}
\label{eq:c_momentum}
    C = \frac{\mvec{\mathcal{L}}\cdot \big[2\widehat{\overline{h}}^2 \widehat{\overline{\mvec{S}}}_\ominus (\mvec{\nabla}\otimes \widehat{\overline{\mvec{u}}})^{\mathrm{T}} \big]}{||2\widehat{\overline{h}}^2 \widehat{\overline{\mvec{S}}}_\ominus (\mvec{\nabla}\otimes \widehat{\overline{\mvec{u}}})^{\mathrm{T}}||^2} = \frac{1}{2}\frac{\mvec{\mathcal{L}}\cdot \mvec{\alpha}}{ ||\mvec{\alpha}||^2},
\end{equation}

\noindent where we have defined $\alpha \equiv \widehat{\overline{h}}^2 \widehat{\overline{\mvec{S}}}_\ominus (\mvec{\nabla}\otimes \widehat{\overline{\mvec{u}}})$.  Using this determined value of $C$, we use equation~(\ref{eq:finaldiffcoeff}) as the diffusivity tensor in the additional flux term from equation~(\ref{eq:realflux}).  Note that $\overline{h}$ is the \textit{resolution scale} and not the radius of compact support of the kernel.  The resolution scale is approximately half of the radius of compact support, or the mean interparticle spacing.

Applying the same procedure to the metal field yields a separate equation for $C_\mathrm{Z}$.  The resolved fluctuations take a similar form to equation~(\ref{eq:resolvedflux}),

\begin{equation}
\label{eq:resolvedmetalflux}
\mvec{\mathcal{L}}_\mathrm{Z} = \widehat{\mvec{\overline{u}}\,\overline{Z}} - \widehat{\mvec{\overline{u}}}\,\widehat{\overline{Z}}.
\end{equation}

\noindent Note that $\mvec{\mathcal{L}}$ is now a vector rather than a rank-2 tensor.  Following the procedure we outline above results in an equation for $C_\mathrm{Z}$

\begin{equation}
\label{eq:c_metals}
C_\mathrm{Z} = \frac{\mvec{\mathcal{L}}_\mathrm{Z}\cdot\big[2\widehat{\overline{h}}^2(\mvec{\nabla}\otimes \widehat{\overline{\mvec{u}}})\cdot\mvec{\nabla}\widehat{\overline{Z}}\big]}{||2\widehat{\overline{h}}^2(\mvec{\nabla}\otimes \widehat{\overline{\mvec{u}}})\cdot\mvec{\nabla}\widehat{\overline{Z}}||^2} = \frac{1}{2} \frac{\mvec{\mathcal{L}}_\mathrm{Z}\cdot\mvec{\beta}}{||\mvec{\beta}||^2},
\end{equation}

\noindent where we have defined $\mvec{\beta} \equiv \widehat{\overline{h}}^2(\mvec{\nabla}\otimes \widehat{\overline{\mvec{u}}})\cdot\mvec{\nabla}\widehat{\overline{Z}}$.

\subsection{Comparison to the Smagorinsky model}
\label{sec:model_comparison}

The gradient model differs from the widely used Smagorinsky model in a subtle yet important way.  Returning to the definition of the sub-grid flux in equation~(\ref{eq:realflux}), the Smagorinsky model represents $\mvec{F}$ as,

\begin{equation}
\mvec{F} = 2 \overline{\rho} (C_\mathrm{s} h)^2 ||\mvec{S}^*|| \mvec{S}^*,
\end{equation}

\noindent where $\mvec{S}^*$ is the trace-free symmetric shear tensor\footnote{$\mvec{S}^* \equiv \mvec{S} - \frac{1}{3}\mathrm{tr}(\mvec{S})\cdot\mvec{I}$}.  The diffusivity tensor is isotropic,

\begin{equation}
\mvec{K}_\mathrm{Smag} \equiv 2 \overline{\rho} (C_\mathrm{s} h)^2 ||\mvec{S}^*|| \, \mvec{I}.
\end{equation}

\noindent and $||\mvec{K}_\mathrm{Smag}|| = 2\overline{\rho} (C_\mathrm{s} h)^2 ||\mvec{S}^*||$ with $C_\mathrm{s} \sim 0.15$.  The constant nature of $C_\mathrm{s}$ implies that the diffusivity scales with the magnitude of the symmetric shear or, more directly, that the strength of turbulent fluctuations is only determined by fluctuations in the shear strength.  That is a good assumption in purely turbulent flows but fails dramatically in laminar shear flows, where the shear is not a good indicator of the presence of turbulence.

One solution to over-diffusion in the Smagorinsky model is to dynamically calculate the coefficient $C_\mathrm{s}$ at simulation time based on the local fluid properties.  We showed in \cite{Rennehan2019} that the dynamic procedure predicts much lower values of $C_\mathrm{s}$ in the majority of our simple hydrodynamical tests.  However, we did not consider the impact of altering the isotropic nature of the diffusivity.

The gradient model has $\mvec{K}_\mathrm{Grad} \propto \mvec{S}_\ominus$ with $||\mvec{K}_\mathrm{Grad}|| = 2\overline{\rho} Ch^2 ||\mvec{S}_\ominus||$.  The constant $C$ is yet to be determined but the direction differs from the dynamic and non-dynamic Smagorinsky models.  The diffusivity itself no longer acts in each direction equally but acts in the direction of the eigenvectors of $\mvec{S}_\ominus$.  However, in a simple incompressible, low Mach number turbulent flows we expect that $||\mvec{K}_\mathrm{Smag}|| \sim ||\mvec{K}_\mathrm{Grad}||$ given that the velocity derivative tensor $\mvec{\nabla}\otimes\mvec{\tilde{u}}$ is approximately isotropic in that regime.   

Our application of the dynamic method \citep{Piomelli1994} to the gradient model simultaneously allows $C = C(\mvec{x}, t)$ and the diffusivity to be anisotropic, $\mvec{K}_\mathrm{Grad} \propto \mvec{S}_\ominus$.  This should be important for any complicated astrophysical flows such as those we investigate in the following sections.

\subsection{Models}
\label{sec:model_list}

\begin{table}
    \centering
    \caption{Turbulence models and parameters.  The dynamic model calculates the model parameters at simulation and, therefore, the values listed below are the forced upper limit.}
    \label{tbl:models_list}
    \begin{tabular}{c|ccc}
        \hline
        Label & Dynamic & (Max.) Parameter Value & Anisotropic \\
        \hline
        \hline
        None & N/A & N/A & N/A \\
        Smag. & \xmark & 0.15 & \xmark \\
        Dyn. Smag. & \checkmark & 0.20 & \xmark  \\
        FIRE & \xmark & 0.05 & \xmark \\
        Grad. & \xmark & 0.22 & \checkmark \\
        Dyn. Grad. & \checkmark & 1.00 & \checkmark \\
        \hline
    \end{tabular}
\end{table}

Table~\ref{tbl:models_list} contains a compact description of our model set.   In all cases where there is a sub-grid turbulence model, we treat both metals and viscosity simultaneously.  There are three categories of models: no sub-grid model (\model{None}), the Smagorinksy model, and the gradient model.  The dynamic procedure allows us to extend the Smagorinsky and gradient models with a model parameter that depends on spatio-temporal coordinates.  Additionally, we test the only other calibration of the Smagorinsky model in the mesh-free finite mass method from the FIRE collaboration \citep{Escala2017}.

For the Smagorinsky model, \model{Smag.}, we use the theoretical value of \smag$\sim0.15$ and limit \smag to $0.20$ for the dynamic Smagorinsky model (\model{Dyn. Smag.}) to avoid numerical instability.  The FIRE calibration of the Smagorinsky model (labelled \model{FIRE}) is \smag$\approx 0.046$ and we adopt \smag$=0.05$ for simplicity. For these three versions of the Smagorinsky model the value of \smag is the same for both metals and eddy-viscosity.

In the new gradient model we use fixed values of $C = 0.22$ and $C_\mathrm{Z} = 0.22$ for the baseline comparison and label these as \model{Grad}.  In our other tests, we use the dynamic procedure outlined in Section~\ref{sec:methods_dynamic_model} and label these tests as \model{Dyn. Grad}.  We derive our fixed values of $C$ and $C_\mathrm{Z}$ from the approximate median value predicted by the dynamic procedure in the driven turbulence tests in Section~\ref{sec:driven_turbulence}.

\section{Homogeneous Turbulence}
\label{sec:driven_turbulence}

Turbulence is ubiquitous in astrophysical flows on a myriad of scales and Mach numbers.  Therefore, in this section, we investigate the impact of the gradient model on the velocity statistics and metal distributions in homogeneous, isotropic, driven turbulence at Mach numbers $\mathcal{M} \in \{0.3, 0.7, 2.1\}$.

For each $\mathcal{M}$, our control (i.e., no sub-grid turbulence model) simulation set comprises $5$ simulations with particle counts $N \in \{64^3, 128^3, 256^3, 512^3, 768^3\}$ within a box of side length $L = 1$ with initial pressure, density, and specific internal energy of $P = 1$, $\rho = 1$, $u = 1000$, respectively\footnote{We use arbitrary code units in all of our hydrodynamical tests.}.  Initially, we place equal mass particles on a uniform grid and then subsequently mix the gas via the prescription of \cite{Schmidt2006}, ported to particle-based simulations in \cite{Price2010a} and subsequently later implemented in \pkg{GADGET} and \pkg{GIZMO} \citep{Bauer2012, Hopkins2015a}.  We list our turbulent driving parameters for \pkg{GIZMO} in Table~\ref{tbl:driven_turbulence} (cf. Table 1 in \citealt{Bauer2012}).  For more details of our approach, see Section 3.1 of \cite{Rennehan2019}.  

\begin{table}
    \centering
    \caption{Parameters for our driven turbulence experiments.  Units are arbitrary code units.}
    \label{tbl:driven_turbulence}
    \begin{tabular}{c|ccccccc}
        \hline
        $\sim\mathcal{M}$ & $\sigma$ & $\Delta t$ & $t_\mathrm{s}$ & $k_\mathrm{min}$ & $k_\mathrm{max}$ & $\tau_\mathrm{mix}$ \\
        \hline
        \hline
        0.3 & 0.014 & 0.005 & 1 & 6.27 & 12.57 & 3.33 \\
        0.7 & 0.045 & 0.005 & 1 & 6.27 & 12.57 & 1.67 \\
        2.1 & 0.21 & 0.005 & 0.5 & 6.27 & 12.57 & 0.50 \\
        \hline
    \end{tabular}
\end{table}

Our interest lies in measuring the impact of (1) the eddy-viscosity model on the velocity power spectra of these driven turbulence volumes and (2) the convergence of metal distribution functions in these volumes.  However, these properties rely on driven turbulence volumes that are in statistical equilibrium.  To gauge whether our simulations are in equilibrium, we define a mixing timescale $\tmix \equiv L / \langle v \rangle$, where $\langle v \rangle$ is the expected average velocity of the particles in each volume.  However, $\langle v \rangle = v_\mathrm{s} \mathcal{M}$, where $v_s = 1$ is the isothermal sound speed of the gas and, additionally, $L = 1$.  Therefore, $\tmix = 1 / \mathcal{M}$.  We evolved each control simulation for several mixing timescales ($\sim 4 \tmix$) to ensure that the gas is in steady-state statistical equilibrium and we confirmed the stability of the Mach number.  The mixing timescale and steady-state Mach numbers are listed in Table~\ref{tbl:driven_turbulence}.  We measure the velocity power spectra following the same method as \cite{Bauer2012}, which is available in the public version of \pkg{GIZMO}. 

We note that there are debates in the literature over the most accurate method to compute the velocity power spectrum in Lagrangian hydrodynamical methods, particularly for classic smoothed particle hydrodynamics \citep{Bauer2012, Price2012}.  The biggest issue is reproducing the correct power on the smallest scales, near the maximum resolution.  \cite{Shi2013} compared several methods and showed that a second-order moving least squares method produces the least error in reproducing the correct velocity power spectra on the smallest scales.  However, it is not clear how this will generalise to the MFM method where the power on smaller scales, as we discuss below, more closely resembles grid-based hydrodynamical methods.   We choose to use the module available in the \pkg{GIZMO} for easy comparison to the turbulence results \cite{Bauer2012} and \cite{Hopkins2015a}, and leave comparison of the different power spectra calculation methods to future work.

It may seem out of place to study low resolutions, such as $64^3$ and $128^3$, in this study when it is clearly possible to study resolutions up to $768^3$.  The ultimate goal of our work is to apply the model to cosmological simulations which have a huge dynamic range and, therefore, low resolutions in individual galaxies.  For example, IllustrisTNG \citep{Pillepich2018} and the RomulusC \citep{Tremmel2018} simulations both have particle mass resolutions of $\sim 10^5$ $\Msun$ for each gas particle, at their best.  Consider that in an $L^*$-galaxy, we expect perhaps $\lesssim10^{10}$ $\Msun$ of hot gas in the circumgalactic medium \citep{Anderson2010}.  At the best resolutions we have today, that gives $\sim 10^5$ particles per $L^*$-halo or $\sim50^3$ particles.  Evidently, contemporary cosmological simulations that capture both hundreds of $\Mpc$ on the large-scale as well as individual galaxies are far off from $\gtrsim 256^3$ per galactic halo.

\subsection{Velocity power spectra}
\label{sec:driven_turbulence_spectra}

A standard measure to determine if eddy viscosity models improve the accuracy of hydrodynamical simulations is whether the velocity power spectra reproduces the theoretically predicted Kolmogorov scaling, $E(k) \sim k^{-5/3}$.  That scaling holds for incompressible, low Mach number turbulence ($\mathcal{M} \lesssim 1$) but is shallower than the apparent scaling in supersonic turbulence, $E(k) \sim k^{-2}$ \citep{Federrath2013a}.  In physical turbulence, the dissipation scale is demarcated by a sharp decline from the Kolmogorov slope on the smallest scales.  In simulations, the resolution scale forces dissipation to occur on a much larger scale than what would occur in nature as the physical dissipation scale is unresolved.  If the numerical viscosity of a hydrodynamical method cannot rapidly dissipate that energy, there will be a build-up of kinetic energy near the resolution scale that causes an unphysical representation of turbulence.

Eddy-viscosity models introduce additional dissipation in the gas by accounting for the unrepresented scales in the flow, or equivalently by minimizing the error from the missing terms in the equations of motion.  The build-up of kinetic energy is usually observed as a "bump" in the velocity power spectra where there is artificial correlation in the velocities on small scales.  However, before we discuss the impact of eddy viscosity models on the power spectrum we must first test the convergence of the mesh-free finite mass (MFM) method in simulations without eddy viscosity.

\begin{figure}
    \centering
    \includegraphics[width=\columnwidth]{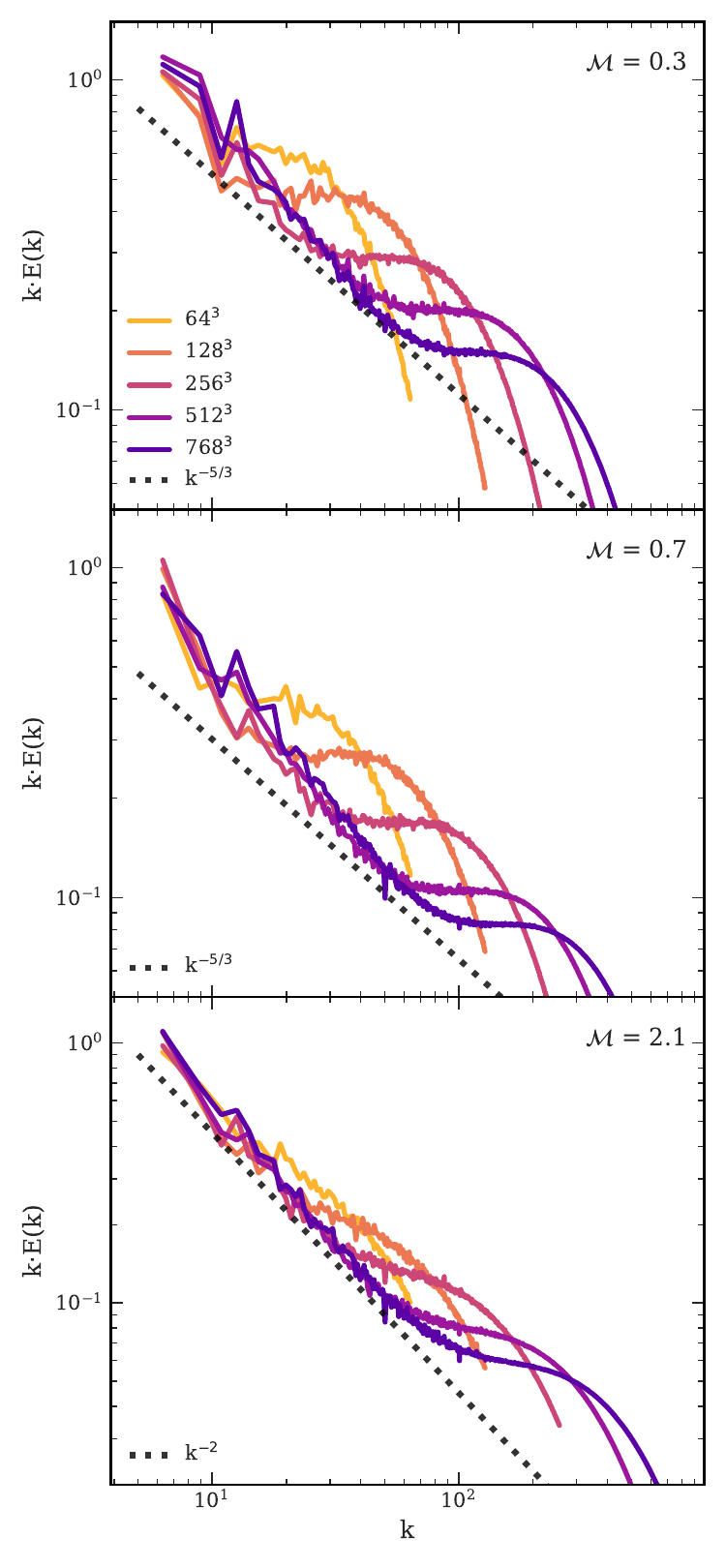}
    \caption{The velocity power spectra of our simulations with no sub-grid eddy viscosity model at three Mach numbers from top to bottom, respectively: $\mathcal{M} = 0.3$, $0.7$, and $2.1$.  We compensate the power spectra by $k$ for easy comparison with \citealt{Bauer2012}.  The coloured lines show resolutions $64^3$, $128^3$, $256^3$, $512^3$, and $768^3$ from lightest to darkest, respectively.  In each panel, we show the predicted scaling with a dotted line of arbitrary normalisation.  Only in the highest Mach number case do we see many of the scales in the inertial range represented and clear convergence of the hydrodynamical method.}
    \label{fig:turb_power_convergence}
\end{figure}

Fig.~\ref{fig:turb_power_convergence} shows the velocity power spectra for our set of simulations with particle counts $64^3$, $128^3$, $256^3$, $512^3$, and $768^3$ coloured by lines from lightest to darkest, respectively.  The power spectra are compensated by $k$ for easy comparison to \cite{Bauer2012}.  The panels are ordered from lowest to highest Mach number from top to bottom --- $\mathcal{M} \sim 0.3$, $0.7$ and $2.1$, respectively.  In each panel, the dotted line represents the predicted scaling but at an arbitrary normalisation. 

From the top panel of Fig.~\ref{fig:turb_power_convergence}, it is apparent that $64^3$ and $128^3$ do not faithfully represent a turbulent gas as they are dominated by the bump.  More precisely, the $E(k)$ scaling is much too shallow compared to Kolmogorov turbulence for a wide range of $k$.  Our $256^3$ simulation shows an inkling of the inertial range scaling but is slightly too steep below $k \lesssim 20$ and dominated by the bump at $k \gtrsim 30$.  As we move up in resolution the inertial range only grows slightly.  At our highest resolution, the inertial range spans $k\sim 40$ to $k\sim 60$ and the bump dominates the small scales.  We skip discussion of the middle panel as the results are qualitatively equivalent between $\mathcal{M} \sim 0.3$ and $0.7$.

In the bottom panel of Fig.~\ref{fig:turb_power_convergence}, we show the compensated power spectra for supersonic turbulence at $\mathcal{M} \sim 2.1$.  It is immediately evident that the simulations converge much more rapidly to the proper scaling than in subsonic turbulence.  Still, at $64^3$ resolution the power spectra is dominated by the build-up of kinetic energy near the resolution scale, with the inertial range only beginning to appear at $128^3$ resolution.  At our highest resolution, $768^3$, there is arguably an entire order-of-magnitude resolved in the inertial range before the build-up of kinetic energy dominates at the smallest scales.

\begin{figure*}
    \centering
    \includegraphics[scale=0.7]{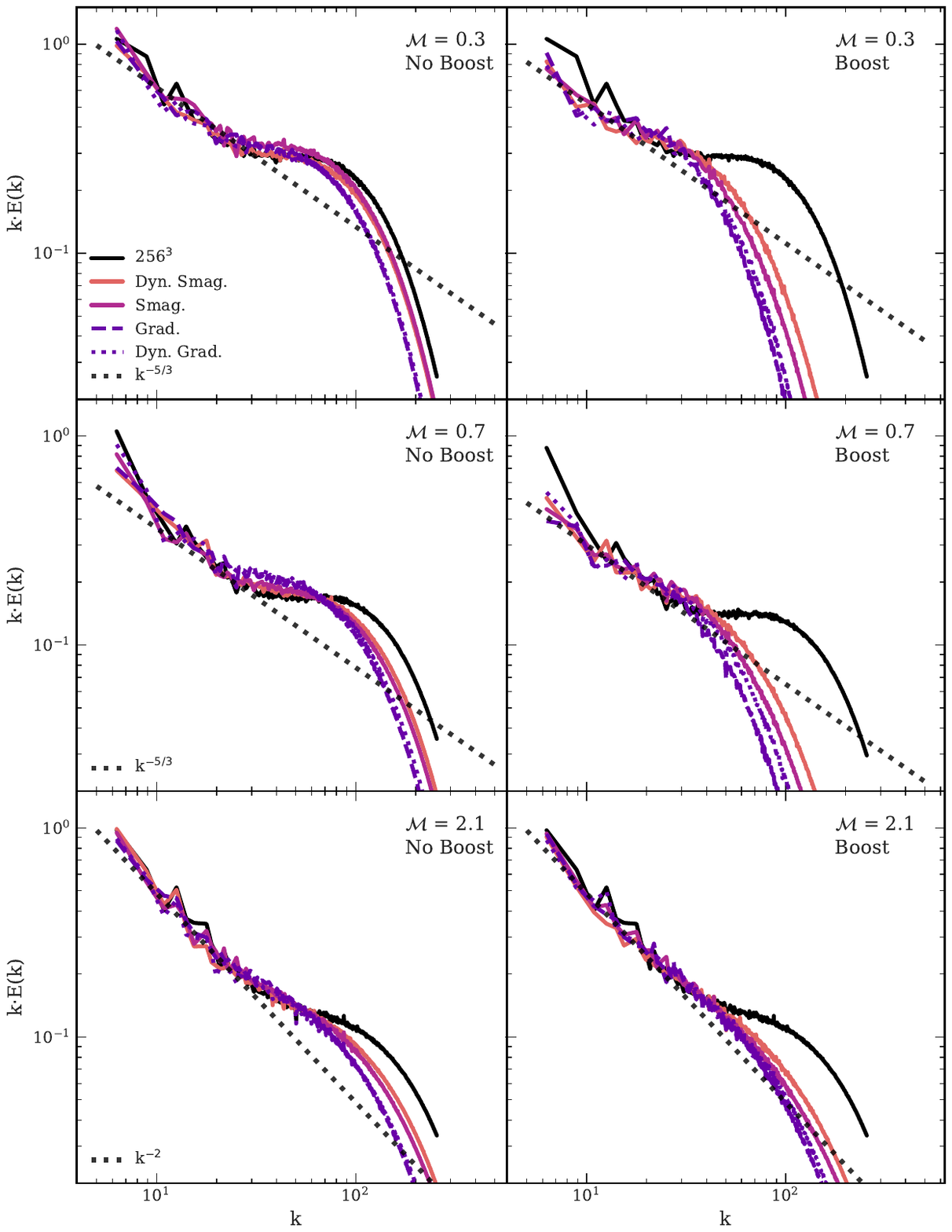}
    \caption{The velocity power spectrum of our turbulence volumes with various eddy viscosity models.  Rows are Mach numbers $0.3$, $0.7$, and $2.1$ from top to bottom, respectively.  The left column has no additional increase in dissipation whereas the right column has an order-of-magnitude boost in dissipation on particles with $\mathcal{M} < 1$.  The black curve shows the $256^3$ simulation with no eddy viscosity model.  The other solid lines show the \model{Smag.} and \model{Dyn. Smag.} models, and the dashed and dotted lines show the \model{Grad.} and \model{Dyn. Grad.} models, respectively.  While all of the models improve the inertial scaling, an additional boost factor ($\gamma \sim 10$) for subsonic particles is required to reproduce the proper scaling at all Mach numbers.}
    \label{fig:turb_power_eddy}
\end{figure*}

Evidently, sub-grid eddy viscosity models are required for the mesh-free finite mass (MFM) method at any resolutions that may be used in astrophysical environments.  We show the impact of our eddy viscosity models in Fig.~\ref{fig:turb_power_eddy}.  The left column shows the compensated power spectra, $E(k)$, as a function of wavenumber $k$.  The rows represent the same three Mach numbers $\mathcal{M} = 0.3$, $0.7$, and $2.1$ from top to bottom, respectively.  The dotted black line shows the Kolmogorov scaling at each Mach number.  All of the simulations were run at $256^3$ resolution since that is the resolution where, with no eddy viscosity model, we begin to see an extended inertial range and distinguish the kinetic energy "bump".  The coloured curves show the eddy viscosity models, shaded from lightest to darkest: \model{Dyn. Smag} (solid salmon), \model{Smag.} (solid magenta), \model{Dyn. Grad.} (dotted purple), and \model{Grad.} (dashed purple).  To explain the right column, we must first explain the results in the left column.

The left column of Fig.~\ref{fig:turb_power_eddy} shows the velocity power spectra of the turbulent gas in our simulated volumes with the same models presented in Section~\ref{sec:methods}.  The black curve shows the control experiment at $256^3$ resolution, i.e. the simulation with no eddy viscosity model and only numerical dissipation as in Fig.~\ref{fig:turb_power_convergence}.  All Mach numbers show the same trend: the eddy viscosity models have little impact on reducing the build-up of kinetic energy at small scales. Especially important is that the sub-sonic ($\mathcal{M} \lesssim 1$) simulations are much less improved than the supersonic case.  However, the new gradient model variants, \model{Grad.} and \model{Dyn. Grad.}, dissipate slightly more rapidly and allow for a steeper slope closer to the Kolmogorov scaling.

We did not expect that all of the eddy viscosity implementations would fail to reduce the kinetic energy build-up \textit{a priori}.  The fact that there is not enough dissipation suggests that some physical property in the diffusion tensor was assigned incorrectly.  As we outlined in Section~\ref{sec:model_comparison}, the diffusion strength for the Smagorinsky and the gradient model classes should effectively scale with each other ($||\mvec{K}_\mathrm{Smag}|| \sim ||\mvec{K}_\mathrm{Grad}||$) in isotropic, homogeneous turbulence.  Therefore, in both classes of models, there are only two physically-motivated quantities that control the strength of diffusion: the length-scale $h$ and the velocity tensor $\mvec{\nabla} \otimes \mvec{u}$.  The velocity tensor should not be the issue since it has been verified through the hydrodynamical tests in \cite{Hopkins2015a} and would cause the MFM method to fail drastically if the velocity gradients were incorrect.  That leaves $h$ as the issue, suggesting that our estimate of the scale over which the eddy viscosity interactions propagate is underestimated\footnote{Recall that our definition of $h$ is already twice as large as \cite{Hopkins2016b}, leading to $4$ times as much dissipation and mixing.}.  Therefore, we introduce a boost factor $\gamma$ to the diffusion tensor $||\mvec{K}||$ (i.e., $||\mvec{K}|| \rightarrow \gamma ||\mvec{K}||$) in order to get a more reasonable scaling in the inertial range.

We determined the boost factor by running a series of driven turbulence tests with discrete $\gamma$ from $\gamma = 1$ to $\gamma = 100$.  We did not perform a quantitative fit to the Kolmogorov scaling as our interest is in the approximate offset required to improve the inertial scaling and the exact $\gamma$ is unimportant for the statistics of the flow.  We additionally found that we only need to correct the diffusion strength in particles that are subsonic, $\mathcal{M} < 1$, with the Mach number for each particle derived from the current velocity of the particle divided by its thermal sound speed.

The right column of Fig.~\ref{fig:turb_power_eddy} shows the velocity power spectra of our simulations with a dissipation boost factor of $\gamma \sim 10$ on only the subsonic particles in each simulation.  At all Mach numbers the additional dissipation causes the kinetic energy to convert into thermal energy much more readily, causing the disappearance of the additional power at small scales, $k \gtrsim 40$.  The gradient models \model{Grad.} and \model{Dyn. Grad.} perform better than the Smagorinksy variants, but only slightly.  That is expected in isotropic homogeneous turbulence since the dissipation strength is effectively the same, and confirms that our implementation of the gradient model is a successful eddy viscosity model.

It is very important that we emphasise the results we observe in Fig.~\ref{fig:turb_power_convergence} are very similar to those found in \cite{Bauer2012} and \cite{Hopkins2015a} for the moving-mesh method (MM; as implemented in \pkg{AREPO}) and the mesh-free finite mass method (MFM; as implemented in \pkg{GIZMO}), respectively.  The only difference is that we show results for much higher resolutions where we begin to see an extended inertial range.  Both the MFM and MM hydrodynamical methods produce the same kinetic energy bump that exists in grid-based methods and, therefore, we would expect an eddy viscosity model to also solve the problem in the MM method, although we do not test that in this work.  To reiterate, it is necessary that \textit{all} hydrodynamical simulations resolve the inertial range combined with an immediate sharp drop-off in power, or they are not reproducing what we physically observe as turbulence below a certain scale, where the build-up of kinetic energy begins to dominate.

\subsection{Metal mixing}
\label{sec:driven_turbulence_metals}

After each simulation reached $\sim 4 \tmix$, we treated each steady-state volume as new initial conditions for our metal mixing study.  In each volume, we gave the densest $50\%$ of particles a metal mass fraction of $Z = 1$ while keeping the rest $Z = 0$.  The metals in our simulations act as passive scalars and have no impact on the flow properties.  We will test the model with more realistic metal distributions in Section~\ref{sec:cosmological}. 

First, we must determine if our simulations converge toward a solution for the metal distribution as we increase resolution.  We ran each of the metal enriched volumes for an additional $4\tmix$ to sample a wide variety of metal distribution states.  We expect \textit{a priori} that by $\sim 2\tmix$ the metal-enriched particles should be scattered approximately homogeneously since a particle with the typical velocity $\langle v \rangle$ should have crossed the volume twice in that time.  Although that is true for all resolutions, how can we compare each resolution on equal footing after it has reached equilibrium? 

The appropriate comparison involves smoothing the spatial distribution of metals on the same scale in all of our simulations.  The main assumption is that our simulations with particle counts $\geqslant 128^3$ contain more accurate information on the scales equivalent to our $64^3$ simulations.  Equivalently, if we \textit{degrade} the resolution of the highest resolution simulations to the lowest resolution, we should hope to obtain a result similar to the lowest resolution simulation.  To degrade the resolution for each simulation, we first kernel-weight the particle data onto a grid with resolution twice as fine as minimum smoothing length in the simulation, $\Delta x_\mathrm{sim,i}$.  Next, we smooth the grid data on a physical scale equivalent to our $64^3$ simulation using a uniform top-hat filter\footnote{Specifically, we use the \pkg{uniform\_filter} function from the \pkg{scipy} package in Python, with periodicity enabled.} with width $w = \Delta x_\mathrm{low} / \Delta x_\mathrm{sim,i}$, where $\Delta x_\mathrm{low}$ is always $\Delta x_\mathrm{low} \equiv 1/64$ since our box has length $L = 1$.

\begin{figure}
    \centering
    \includegraphics[scale=0.64]{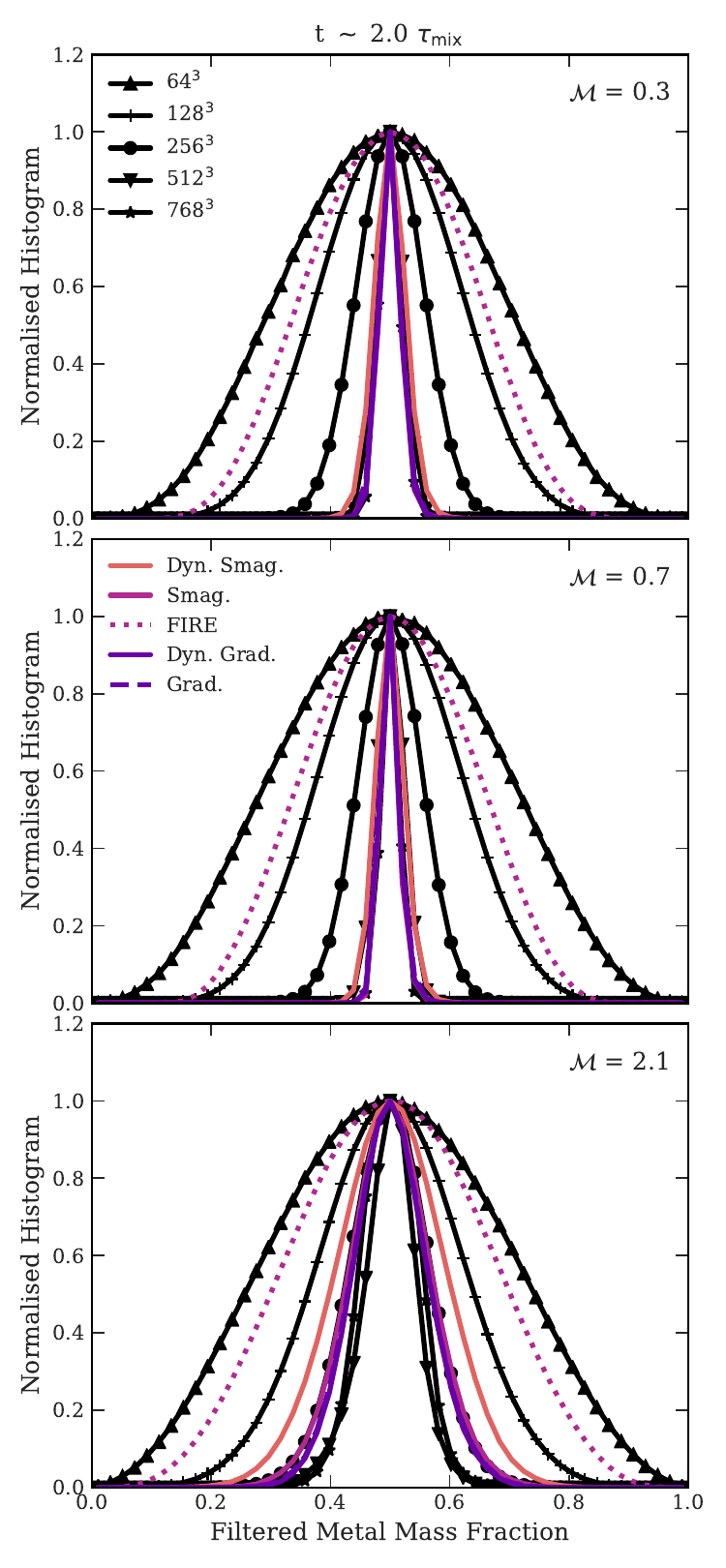}
    \caption{The normalised histograms of the filtered metal mass fraction field from the simulations in Section~\ref{sec:driven_turbulence_metals}.  The panels show Mach numbers $\mathcal{M} = 0.3$, $0.7$, and $2.1$ from top to bottom, respectively after two mixing timescales.  The black curves show the simulations with no sub-grid metal mixing.  The coloured curves show the simulations at $64^3$ resolution with a sub-grid metal mixing model given by the label.  All of the sub-grid metal mixing models show improvement except for \model{FIRE} which lags due to the lower calibration coefficient.}
    \label{fig:turb_metal_dists}
\end{figure}

Fig.~\ref{fig:turb_metal_dists} shows the normalised histograms of the filtered metal field.  The panels are Mach numbers $\mathcal{M} = 0.3$, $0.7$, and $2.1$ from top to bottom, respectively.  The black curves with markers show the convergence of the filtered metal field for resolutions $64^3$, $128^3$, $256^3$, $512^3$, and $768^3$.  The coloured lines show, from lightest to darkest: \model{Dyn. Smag} (solid salmon), \model{Smag.} (solid magenta), \model{FIRE} (dotted magenta), \model{Dyn. Grad.} (dashed purple), and \model{Grad.} (solid purple) at $64^3$ resolution.  We obtained all of the information for this figure after approximately two mixing timescales, $t \sim 2\, \tau_\mathrm{mix}$, where $\tau_\mathrm{mix} = 1 / \mathcal{M}$.  All of the sub-grid models except the \model{FIRE} calibration predict a more accurate large-scale metal distribution, at lower resolution.

\begin{figure}
    \centering
    \includegraphics[width=\columnwidth]{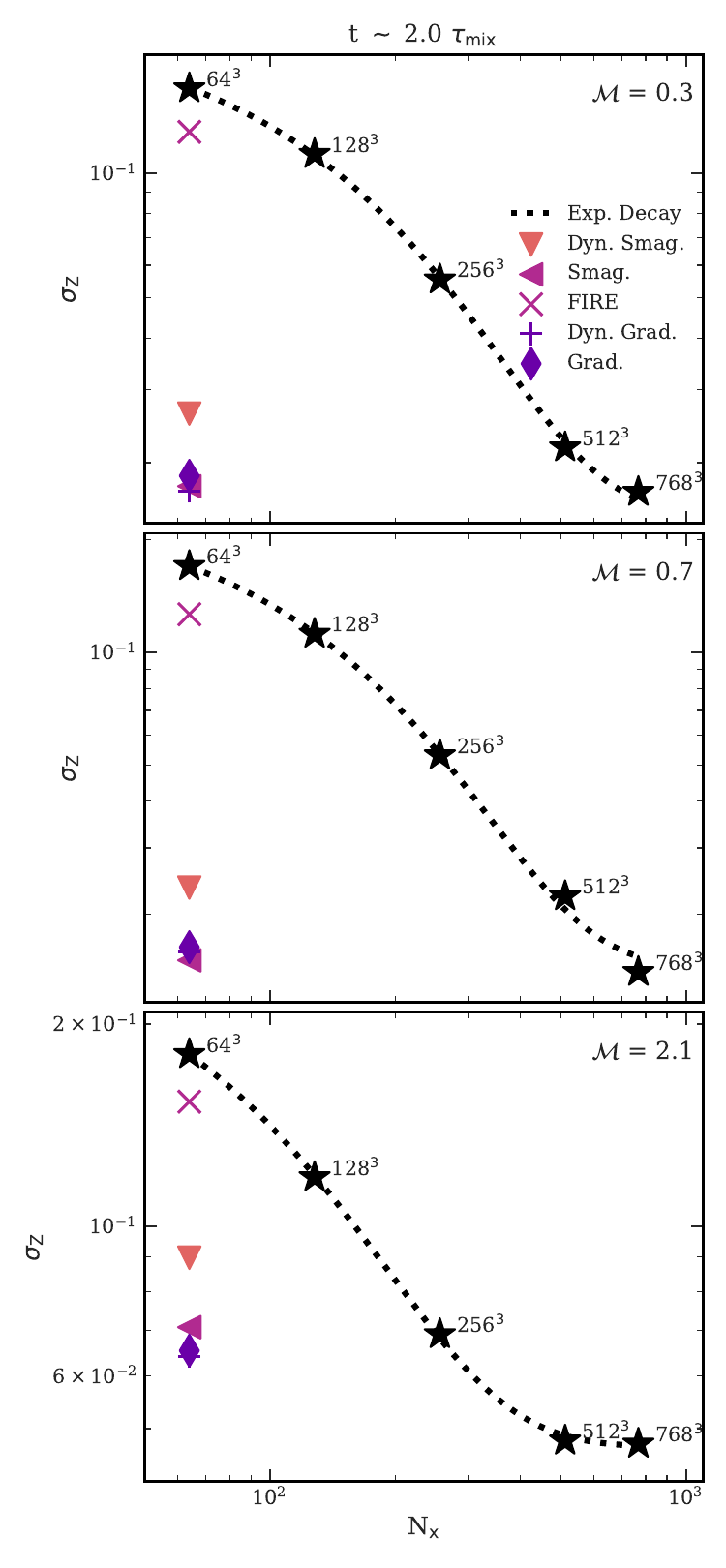}
    \caption{The standard deviation of the Gaussian fit to the metal distributions in our driven turbulence experiments as a function of resolution, $N_\mathrm{x}$.  Mach numbers are $\mathcal{M} = 0.3$, $0.7$, and $2.1$ from top to bottom, respectively.  The stars show the simulations with no sub-grid metal mixing model at resolutions given by the number associated with each point.  The simulations at $64^3$ resolution with sub-grid metal mixing models are given by the figure labels.  The dotted line shows an exponential fit to the no metal mixing model simulations.  Clearly, the gradient model allows us, at $64^3$  resolution, to reproduce the spatial redistribution of metals due to turbulence at a level better than $4 - 8$ times the resolution.}
    \label{fig:turb_convergence}
\end{figure}

Fig.~\ref{fig:turb_convergence} shows the standard deviation ($\sigma_\mathrm{Z}$) of the smoothed metal distribution as a function of $N_\mathrm{x} = \{64, 128, 256, 512, 768\}$ in our simulations at $t \sim 2 \tmix$, for Mach numbers $0.3$, $0.7$, and $2.1$ from top to bottom, respectively.  The stars correspond to the simulations without a sub-grid metal mixing model at the resolution given by their labels.  The remaining symbols in the legend correspond to the simulations at $64^3$ with a sub-grid metal mixing model (see Table~\ref{tbl:models_list} for a description).  The dotted line shows an exponential decay fit to test convergence in the simulations without metal mixing.  We expect decreasing $\sigma_\mathrm{Z}$ with increasing resolution since hydrodynamical mixing is more resolved and the metal value in each grid cell approaches the mean.

At $\mathcal{M} \sim 0.3$ in the top panel of Fig.~\ref{fig:turb_convergence}, $\sigma_\mathrm{Z}$  follows an exponentially decreasing trend with resolution in the simulations without metal mixing, as expected.  However, there is not evidence for strong convergence at our highest resolution --- although it appears the curve is beginning to flatten.  The inverted triangle shows the results for the \model{Dyn. Smag.} model and the left-pointing triangle shows the result for the \model{Smag.} model.  Both the dynamic and standard Smagorinsky models predict a more reasonable $\sigma_\mathrm{Z}$, reproducing a metal distribution closer to a resolution of $512^3$.  The \model{FIRE} calibration of the Smagorinsky model, marked as \xmark, shows little improvement in $\sigma_\mathrm{Z}$; the result is effectively equivalent to having no model at all.  The \model{Dyn. Grad.} and \model{Grad.} models are marked by a plus sign and diamond, respectively.  The gradient model obviously improves $\sigma_\mathrm{Z}$ and can, at $64^3$ resolution, also reproduce a metal distribution equivalent to a resolution $512^3$.  The trends are equivalent for $\mathcal{M} \sim 0.7$ turbulence, so we  continue to the next panel.

The bottom panel of Fig.~\ref{fig:turb_convergence} shows how the sub-grid metal mixing models impact supersonic turbulence at $\mathcal{M} \sim 2.1$.  There is much better convergence of the metal distribution at $t \sim 2\tmix$ than in subsonic turbulence at this time.  The qualitative trend remains the same for the $64^3$ simulations with sub-grid metal mixing: both the gradient models and Smagorinsky models perform equally as well.  However, the width $\sigma_\mathrm{Z}$ of the distributions are much wider in $\mathcal{M} \sim 2.1$ turbulence.  With the exception of the \model{FIRE} calibration, the other metal mixing models at $64^3$ resolution predict metal distributions similar to $256^3$.

Our turbulence tests with metals demonstrate that sub-grid metal mixing models are necessary in the mesh-free finite mass (MFM) method if one desires more accurate metal distributions.  The method we provided for calibrating the metal mixing models is important and, we argue, must be investigated whenever one implements a new metal mixing model into \textit{any} hydrodynamics solver.  Particularly, one must calibrate the model parameter at the resolution they desire for their mixing model if a dynamic procedure is not applied.  That is an important point: the dynamic procedure (\pkg{Dyn. Smag.} and \pkg{Dyn. Grad.}) allows us to approximate the calibrated model parameter for the corresponding model (\pkg{Smag.} and \pkg{Grad.}) without carrying out the calibration.  However, the true power of the dynamic procedure is in simulations with mixtures of non-turbulent and turbulent gas at various Mach numbers.  In those complex environments the dynamic procedure automatically adjusts the model parameter and, as we showed in \cite{Rennehan2019} for the \pkg{Dyn. Smag.} model, drastically alters the resulting metal distributions. 

As expected, in pure homogeneous, isotropic turbulence all of the sub-grid metal mixing models improved the accuracy of the metal distributions when using a proper calibration.  That is expected in the homogeneous, isotropic case since all of the models effectively act in the same way, on average.  The power of the dynamic procedure, and the new anisotropic model is in cosmological simulations where many complex flows interact.

\subsection{Applicability to other hydrodynamical methods}
\label{sec:driven_turb_hydro}

While our interest lies in the mesh-free finite mass (MFM) method, the models in Section~\ref{sec:methods} and experiments in Section~\ref{sec:driven_turbulence} are applicable to other hydrodynamical solvers.  Specifically, our results extrapolate with minor modification to grid-based hydrodynamics.  Only slight changes to the filtering method must be implemented, as outlined in \cite{Schmidt2015}.  More care must be taken when applying the model to smoothed particle hydrodynamics (SPH). However, as we outline below, there is much broader applicability to the moving-mesh (MM) method.

First we consider eddy viscosity.  In SPH, it is well-established that there is a \textit{deficit} in power near the resolution scale rather than a build-up of kinetic energy \citep{Bauer2012, Price2012, Hopkins2013, Hopkins2015a}.  That fact suggests that SPH reproduces turbulence better than the MFM or MM methods, but produces results at a much lower effective resolution.  The lack of power, rather than the overabundance of power, implies that an eddy viscosity model would only further degrade the resolution of SPH results and not improve the inertial scaling.  Therefore, we do not recommend eddy viscosity models for SPH but rather the work of \cite{DiMascio2017}, who recently provided an SPH equivalent. 

For the MM method, the build-up of kinetic energy at the resolution scale is present \citep{Bauer2012} and of equivalent magnitude to our results in the MFM method.  Therefore, we recommend investigation into eddy viscosity models for the MM method as they could improve the inertial scaling.  In terms of implementation, all of the derivations in Section~\ref{sec:methods} apply to the MM method.

Metal mixing using the Smagorinsky model has been studied in cosmological simulations involving SPH but widely ignored in the MM method.  However, no calibration technique has been provided by the community for SPH and the calibrations usually follow the theoretical value for the Smagorinsky model (e.g., \citealt{Shen2010, Williamson2016a}) or calibrations that depend on sub-grid astrophysics models (e.g., \citealt{Wadsley2017, Escala2017}).  The metal mixing calibration technique in Section~\ref{sec:driven_turbulence} is completely applicable to SPH since metals are treated equivalently to the MFM method, and they are both constant mass methods.  Additionally, our calibration technique does not depend on the uncertainties within astrophysical sub-grid models --- only pure hydrodynamics.

For the MM method, all of Section~\ref{sec:methods} is applicable for metal mixing since the MM method relies on transport equations such as equation~(\ref{eq:momentumtransport}) for advecting metals throughout the fluid \citep{Springel2010}.  In fact, \cite{Balarac2013} find that the gradient model improves the inertial scaling in the power in the metal field through identical transport equations.  However, convergence tests such as those in Fig.~\ref{fig:turb_convergence} are necessary in order to determine whether they are truly required, or if numerical dissipation is adequate.

\section{Cosmological Simulations}
\label{sec:cosmological}

Understanding the evolution of galaxies is a complex enterprise involving highly non-linear coupled physical processes.  Not only do stellar feedback and active galactic nuclei produce powerful outflows that drive turbulence locally in the interstellar media of galaxies, but also in the gas reservoirs surrounding galaxies.  Turbulence also appears through the Kelvin-Helmholtz instability in ram pressure stripping of galaxies moving through a hot medium, and through the stellar winds from stars making their way out from the galaxy into the circumgalactic medium. 

The physical processes above occur on spatial scales much smaller than currently possible to resolve in the average contemporary cosmological simulation.  For that reason, the majority of astrophysics in cosmological simulations are encoded into parametrised sub-grid models that use the information on the largest scales to predict what occurs below the resolution of the simulation.  The resulting calculations usually indicate how much mass and energy should be injected (or removed) from the large-scale gas and stellar components.  However, there is no one correct way to approximate the astrophysics on the sub-grid scale since it highly depends on the maximum possible resolution, hydrodynamical method, as well as other complex numerical effects.  All of the issues with numerics and missing physics usually ends up in one or more tunable free parameters in the model. 

Assuming such a sub-grid astrophysical model is developed, how do we know that it is correct?  Or, at least approximating reality? Normally, one or more trusted astronomical observation is used to test the validity of all of the sub-grid astrophysics that may exist.  A common example would be the galaxy stellar mass function, or the $\mathrm{M}_\mathrm{BH}$ - $\sigma_\mathrm{*}$ relationship that links supermassive black hole masses to the stellar masses of their host galaxies.  However, two different hydrodynamical methods may provide different parameter values for the same sub-grid astrophysical models.  Additionally, there may be two completely different approaches to modelling the same physical phenomenon with no clear mapping between free parameters!  Calibrating sub-grid astrophysical models is obviously a complicated endeavour and must be built on a strong hydrodynamics base.  How can we begin to trust that our understanding of the astrophysics of small scale approximate reality if the hydrodynamics, as we showed in Section~\ref{sec:driven_turbulence}, does not reproduce reality?

Our goal is in determining whether the converged and separately calibrated sub-grid turbulence models we presented in Section~\ref{sec:methods} have any significant impact over the sub-grid astrophysical models that we use in large-volume cosmological simulations.  As a first step, we only investigate the broad, qualitative impact on the gas properties in gaseous halos in a single set of sub-grid astrophysical models.  We stress that we do not intend to reproduce the full galaxy population in a calibrated and predictive sense.  Additionally, we note that more testing is required across the all of sub-grid astrophysical models that exist in the literature, as the turbulent mixing models may interact in unexpected ways due to the non-linearity of the problem.

\subsection{The simulations}
\label{sec:cosmo_simba}

For our comparison, we choose to use the \pkg{SIMBA} galaxy formation model.  \pkg{SIMBA} includes robust sub-grid models of star formation, cooling, stellar feedback, chemical enrichment, active galactic nuclei feedback, and dust evolution --- all evolved in concert with the mesh-free mass method (MFM) \citep{Dave2016c, Dave2019}.  For this study, we implemented the \pkg{SIMBA} models into the public version of \pkg{GIZMO} as described in \cite{Dave2019} and we point the interested reader to that study for the details of the sub-grid models.  We follow the approach of \cite{Schaye2014} and calibrate our implementation of the \pkg{SIMBA} model only to the galaxy stellar mass function and the $M_\mathrm{BH}$-$M_*$ relationship at $z = 0$ for the purposes of this study.

\begin{table}
    \centering
    \caption{Cosmological parameters and simulation information for our simulation set. Our parameters follow \citealt{Dave2019} except begin at a higher redshift.}
    \label{tbl:cosmological_simulations}
    \begin{tabular}{l|l}
        \hline
        Cosmological Parameters  \\
        \hline
        $\Omega_\mathrm{m,0}$ & $0.3$ \\
        $\Omega_\mathrm{\Lambda,0}$ & $0.7$ \\
        $\Omega_\mathrm{b,0}$ & $0.048$ \\
        $h$ & $0.68$\\
        $\sigma_\mathrm{8}$ & $0.82$\\
        $n_\mathrm{spec}$ & $0.97$\\
        \hline
        Simulation Information \\
        \hline
        $z_\mathrm{begin}$ & 249\\
        $N_\mathrm{part}$ & $2\times256^3$ \\
        $L_\mathrm{side}$ & $25$ $\cMpch$ \\
        $m_\mathrm{part,gas}$ & $1.26\times10^{7}$ $\Msunh$\\
        $m_\mathrm{part,dark}$ & $6.88\times10^{7}$ $\Msunh$\\
        $\epsilon_\mathrm{soft,min}$ & $0.5$ $\kpch$\\
        \hline
    \end{tabular}
\end{table}

We run 6 cosmological-scale volumes of side-length $L = 25$ $\cMpch$ ($\sim 37$ $\mathrm{cMpc}$)  in order to compare our various mixing models.  The simulations begin from initial conditions generated with \pkg{MUSIC} \citep{Hahn2011} at a redshift of $z = 249$, with a standard $\Lambda$ Cold Dark Matter cosmology (see Table~\ref{tbl:cosmological_simulations} for values).  The mass resolution in gas and dark matter follows the \pkg{SIMBA} simulations with $m_\mathrm{part,gas} = 1.26\times10^{7}$ $\Msunh$ and $m_\mathrm{part,dark} = 6.88\times10^{7}$ $\Msunh$, respectively.  We use adaptive gravitational softening \citep{Hopkins2015a, Hopkins2018} to compute the softening lengths of all of our particles, and enforce a minimum softening length of $\epsilon_\mathrm{soft,min} = 0.5$ $\kpch$.

\subsection{Global metal mixing}
\label{sec:cosmo_global} 

While sub-grid scale turbulence models maximally impact the smallest scales in a cosmological simulation, their integrated effect impact the properties of the largest scales, such as global metal distribution functions \citep{Shen2010, Escala2017, Rennehan2019}.  Therefore, in this Section, we examine the impact of the gradient model on the metal distribution functions (MDFs) in the circumgalactic medium (CGM) and warm-hot intergalactic medium (WHIM) --- both known to be turbulent environments \citep{Iapichino2011, Iapichino2013, Tumlinson2017}.  

\begin{figure*}
    \centering
    \includegraphics[scale=0.75]{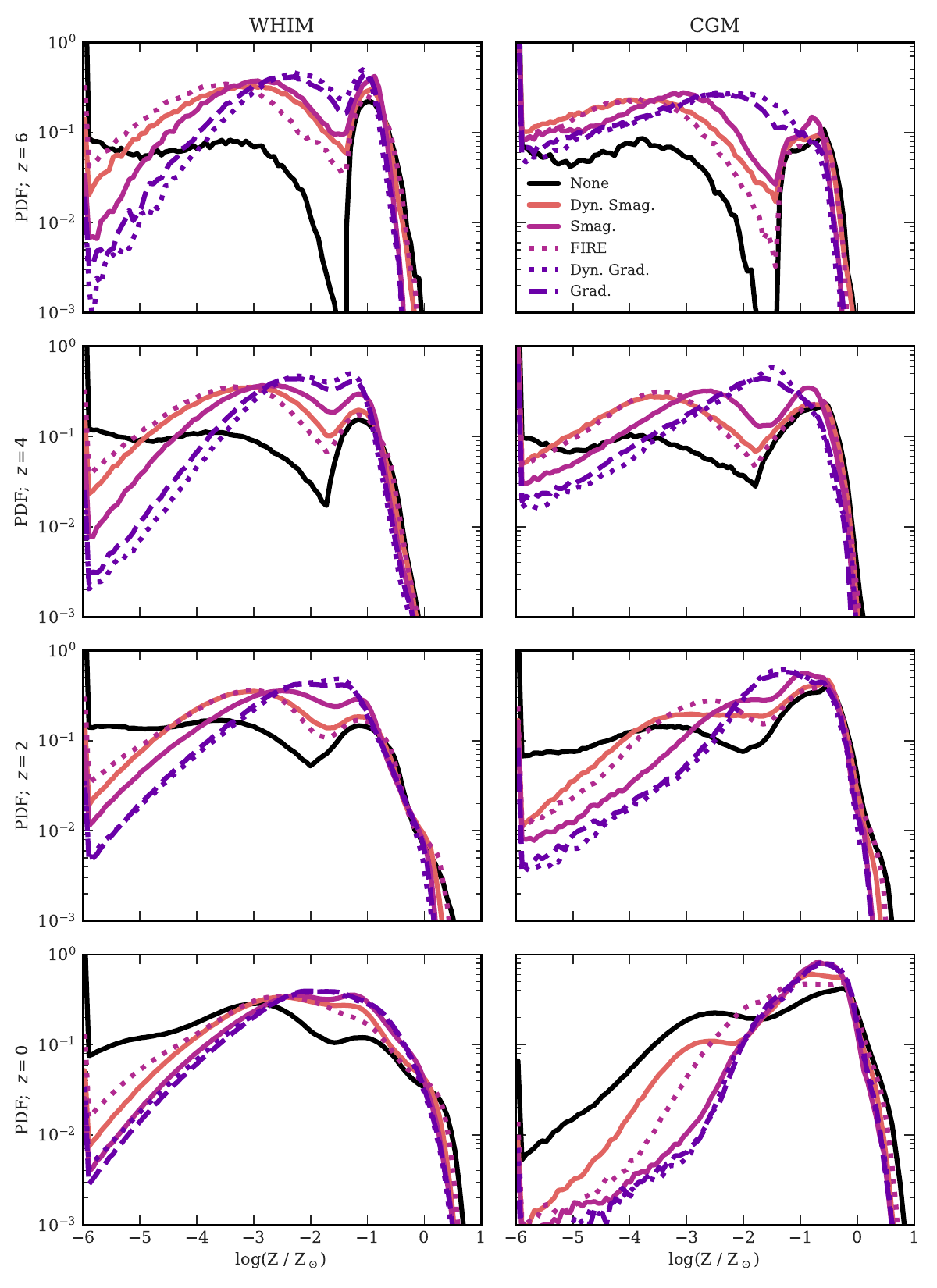}
    \caption{The metal distribution function across global phases: the warm-hot intergalactic medium (WHIM; left column) and the circumgalactic medium (CGM; right column).  Redshifts are given by the rows from top to bottom: $z = 6$, $4$, $2$, and $0$, respectively.  The gradient model variants mix metals much more rapidly in the early stages of the simulation, suggesting that the choice of model will impact studies that focus on enrichment timing of halo gas.}
    \label{fig:phase_metal_dists}
\end{figure*}

Our definition of CGM and WHIM depends on separating gas that is bound to halos from that which is unbound, at a given epoch.  A good estimation comes from \cite{Dave2010},

\begin{equation}
\frac{\rho_\mathrm{bound}(z)}{\Omega_\mathrm{b}(z)\rho_\mathrm{c}(z)} = 6\pi^2 \bigg(1 + 0.4093 \bigg(\frac{1}{\Omega_\mathrm{m}(z)} - 1\bigg)^{0.9052}\bigg) - 1,
\label{eq:bound_density}
\end{equation}

\noindent where $\Omega_\mathrm{b}(z)$ is the baryon fraction as a function of redshift, $\Omega_\mathrm{m}(z)$ the matter fraction, $\rho_{c}(z) = 3(H(z))^2 / (8\pi G)$, and $H(z)$ the redshift-dependent Hubble function.  All gas above $\rho_\mathrm{bound}(z)$ we consider bound to halos, and have confirmed that the approximation holds well.

We define the CGM to be all gas in the volume that is above $\rho_\mathrm{bound}(z)$ in equation~(\ref{eq:bound_density}) and below the star formation density threshold, $\rho_\mathrm{*,crit}   = 4.4\times 10^{-25}$ g cm$^{-3}$, at any temperature.  That includes gas in the intragroup medium of our most massive halos in the $(25 \, \cMpch)^3$ volumes.  The WHIM is all gas that is below $\rho_\mathrm{bound}(z)$ and above a temperature of $T = 10^5$ $\mathrm{K}$.  

Fig.~\ref{fig:phase_metal_dists} shows the metal distribution functions (MDFs) for our two gas phases in columns: WHIM (left) and CGM (right), and at $z = 6$, $4$, $2$, and $0$ in rows from top to bottom, respectively.  These are probability density functions, and were constructed by binning the particle metallicities in the range $10^{-6} < \log(Z/Z_\odot) < 10^{1}$, where $Z_\odot = 0.0134$ \citep{Asplund2009}.  The black curves show the control simulation, \model{None}, with no sub-grid metal mixing.  The coloured curves show the simulations with sub-grid metal mixing and are, from lightest to darkest: \model{Dyn. Smag} (solid salmon), \model{Smag.} (solid magenta), \model{FIRE} (dotted magenta), \model{Dyn. Grad.} (dotted purple), and \model{Grad.} (dashed purple).  See Table~\ref{tbl:models_list} for more details.

First we focus on the WHIM.  At $z = 6$, there are two distinct components across all of our model variants. The peak at $Z \sim 10^{-1}$ $\Zsun$ is the highly enriched interstellar medium (ISM) gas that recently joined the WHIM via stellar winds from the integrated star formation in the early universe.  The lower distribution is gas that has mixed into the WHIM but did not recycle through the ISM, missing the opportunity for further enrichment via supernova feedback.  It is important to note that since there is no mixing in the \model{None} case, when a particle leaves the ISM it \textit{cannot} change its metallicity.   The models that include sub-grid mixing show varying spread in the MDFs, with the \model{FIRE} and \model{Dyn. Smag.} showing the widest spread.  The \model{Dyn. Grad.} and \model{Grad.} models show the tightest distributions, with the two peaks in the distributions seemingly merging at $Z \sim 10^{-1.5}$ $\Zsun$.  The \model{Smag.} model matches the \model{Dyn. Smag.} model at $Z \gtrsim 10^{-3}$ $\Zsun$, but is biased toward higher metallicities below that threshold. 

The next 3 panels in the left column of Fig.~\ref{fig:phase_metal_dists} show the MDFs in the WHIM for $z = 4$, $2$, and $0$, from top to bottom, respectively.  The trend for all models is to approach a singly-peaked distribution as the simulation evolves through cosmic time.  Most of the evolution in the MDFs occurs from $z = 6$ to $z = 2$ after which the distributions are mostly stationary.  The transition from $z = 6$ to $z = 4$ demonstrates how rapidly the WHIM evolves at high redshift, and how each sub-grid metal mixing model impacts the MDFs with different mixing rates.  Specifically, the \model{Dyn. Grad.} and \model{Grad.} models predict similar distributions at $z = 4$, and produce the tightest MDFs compared to all of the other models.  In fact, there is the same trend at $z = 4$ as at $z = 6$ --- the gradient model variants (\model{Dyn. Grad.} and \model{Grad.}) predict tighter MDFs, followed by wider distributions in the Smagorinsky models (\model{Smag.}, \model{Dyn. Smag.}, and \model{FIRE}, respectively).  

There are similar trends in the CGM, as we see in the right column of Fig.~\ref{fig:phase_metal_dists}.  To reiterate, the panels show redshifts $z = 6$, $4$, $2$, and $0$ from top to bottom, respectively.  

At $z = 6$, there is a clear distinction between the distribution at $Z \lesssim 10^{-2}$ $\Zsun$ and $Z \gtrsim 10^{-2}$ $\Zsun$ in the \model{None} case, and the Smagorinsky variants (\model{Smag.}, \model{Dyn. Smag.}, and \model{FIRE}).  Stellar feedback drives the peak at $Z \sim 10^{-1}$ $\Zsun$ similarly to the WHIM at this redshift, whereas the distribution at $Z \lesssim 10^{-2}$ $\Zsun$ is from the very first generations of stars.  By this time the \model{Dyn. Grad.} and \model{Grad.} models have mixed the most rapidly to create a single broad distribution in their MDFs.  All of the models with sub-grid metal mixing have much more gas mass enriched above $Z > 10^{-6}$ $\Zsun$ than the \model{None} case, especially compared to the deficit at $Z \sim 10^{-1.5}$ $\Zsun$ in the \model{None} case.  The Smagorinsky models vary in mixing rate as \model{Smag.}, \model{Dyn. Smag.}, and \model{FIRE}, from fastest to slowest, respectively.  The MDFs in the CGM at redshifts $z = 4$ to $z = 0$ demonstrate the same trends as in the WHIM phase at the same redshifts: the gradient models mix much more rapidly at early stages than the Smagorinsky models.  At $z = 0$ the \model{Dyn. Grad.}, \model{Grad.}, and \model{Smag.} models predict the same distribution in the global CGM phase, whereas the \model{Dyn. Smag.} and \model{FIRE} models predict slightly less enriched gas.

The MDFs in the turbulent WHIM and CGM show the importance of sub-grid metal mixing models in cosmological simulations as well as the importance of model choice.  In all cases we include metal mixing, the MDFs are significantly tighter at all redshifts we measure and significantly tighter for the CGM at $z = 0$.  This is contrary to the study in \cite{Su2016} that found metal mixing to be relatively unimportant on cosmological scales.  However, we find that the \model{FIRE} calibration is much too low to reproduce the correct converged hydrodynamical mixing of metals (see Section~\ref{sec:driven_turbulence_metals}).  With our new calibrations of the Smagorinsky model, \model{Smag.}, and the new gradient models, \model{Dyn. Grad.} and \model{Grad.}, we see significant differences at all redshifts. 

A common theme in theoretical galaxy evolution is that equivalent results between different models at $z = 0$ does not necessarily imply a similar integrated history.  The evolutionary paths for each sub-grid metal mixing model all evolve at slightly different rates as we would expect based on their diffusivities from Section~\ref{sec:methods}.  The lesson from our results is that the metal mixing model choice impacts the early development phases of galaxies rather than the long term equilibrium stages.  At higher redshifts, $z \gtrsim 2$, gas is collapsing to form galaxies, while stellar feedback and supermassive black holes are driving outflows out of the potential wells and forcing turbulence.  Our inclusion of the full diffusion tensor in the \model{Dyn. Grad.} and \model{Grad.} models allows the gas that is compressing from feedback and infall to further mix its metal mass with nearby neighbours, tightening the MDFs.

The Smagorinsky variants also improve the results and allow metal mixing between gas particles but produce broader distributions, notwithstanding the \model{Dyn. Smag.} and \model{Smag.} models showing good matches in the convergence tests of Section~\ref{sec:driven_turbulence_metals}.  This is an important point: the simple turbulence tests in Section~\ref{sec:driven_turbulence_metals} showed agreement between the gradient and Smagorinsky models (except for the lower \model{FIRE} calibration) but now we see disagreement in complex cosmological environments.  Evidently, ignoring the trace of the velocity tensor is not the correct approach for cosmological contexts and we recommend either the \model{Grad.} or \model{Dyn. Grad.} models as we see no difference with the dynamic procedure applied to the constant coefficient gradient case.

\subsection{The impact of eddy viscosity}
\label{sec:cosmo_viscosity}

\begin{figure*}
    \centering
    \includegraphics[scale=0.55]{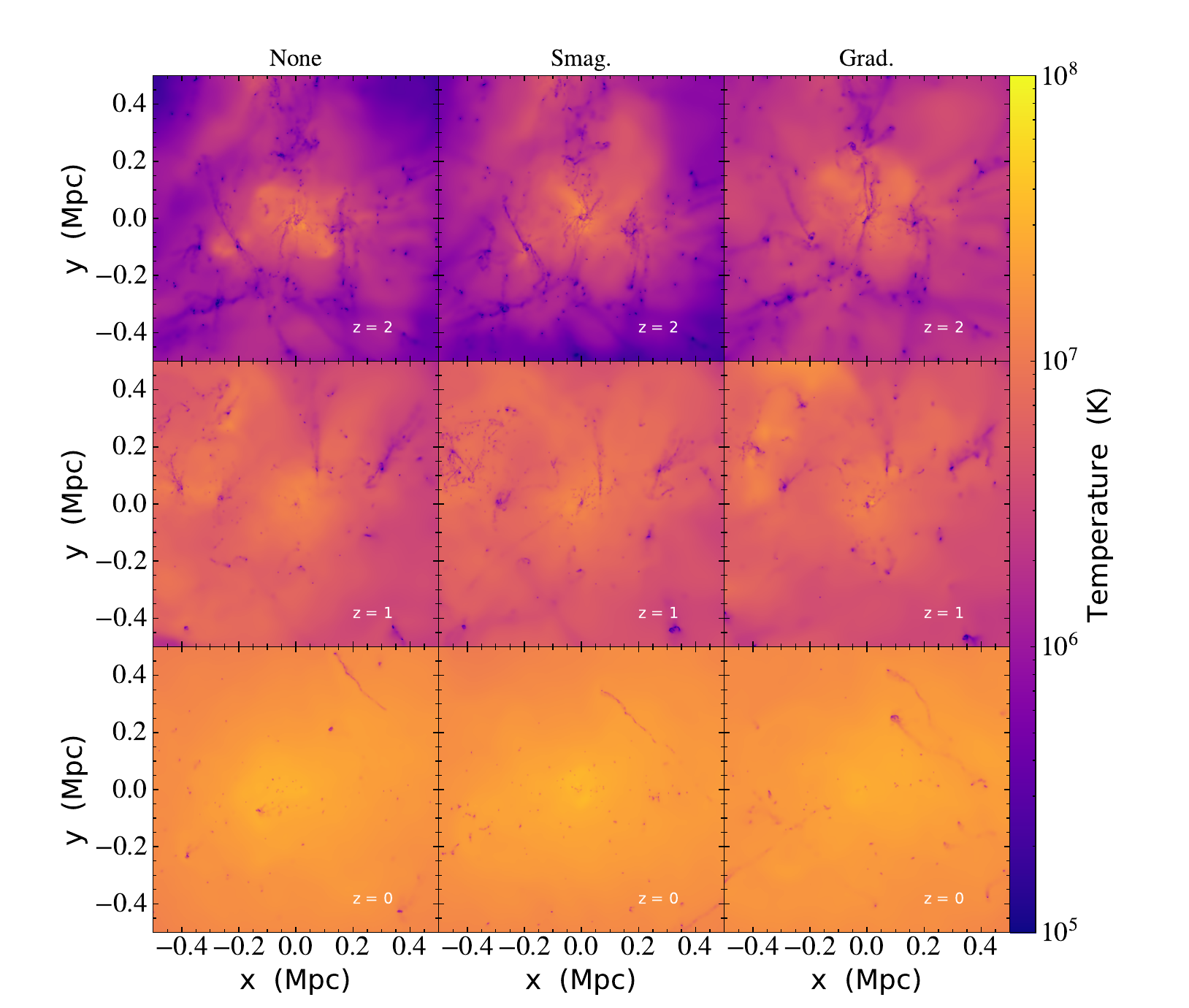}
    \caption{Temperature projections of the most massive halo in three of our cosmological simulations at redshifts $z = 2$, $1$, and $0$ in rows from top to bottom, respectively.  The columns show our \model{None}, \model{Smag.}, and \model{Grad.} models from left to right, respectively.  Each of the panels represents a $1$ $\Mpc$ by $1$ $\Mpc$ (physical) region centred on the most massive galaxy at each redshift.  The \model{Smag.} simulation shows the most small-scale structure at all redshifts, and a smoother distribution of temperature at high-redshift compared to the \model{None} case.  The \model{Grad.} model produces less small-scale structure than any other model, and much more hot gas at high redshift.  Surprisingly, the \model{Grad.} model also produces more extended tails from sub-structure moving through the hot halo at lower redshift, suggesting it may impact future studies of jellyfish galaxies.}
    \label{fig:temperature_proj}
\end{figure*}

\begin{figure*}
    \centering
    \includegraphics[scale=0.8]{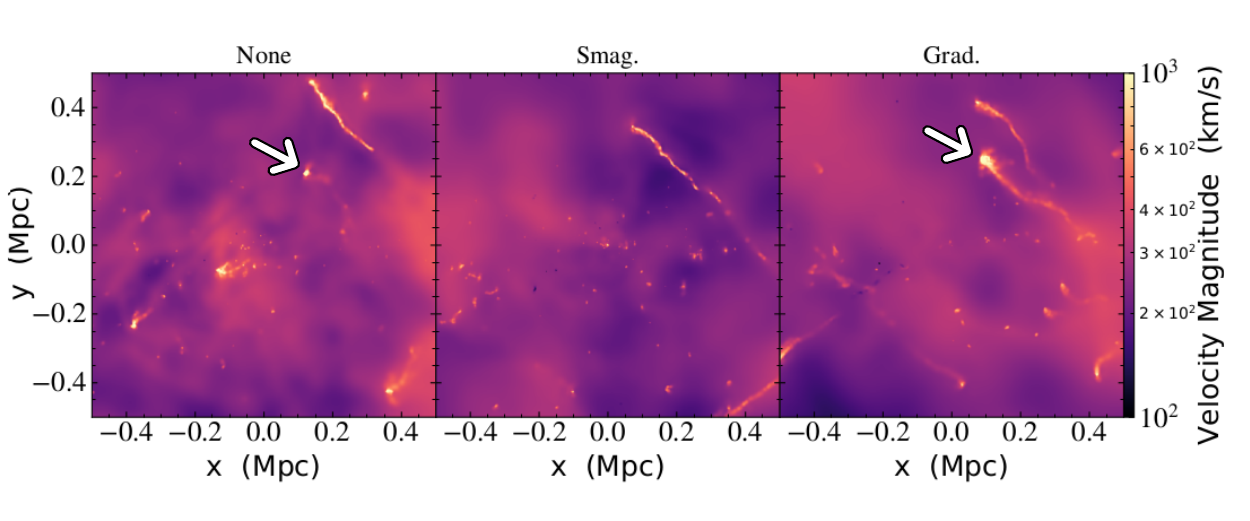}
    \caption{Velocity magnitude projections of the most massive halo at $z = 0$ in our cosmological simulations.  From left to right we show the \model{None}, \model{Smag.}, and \model{Grad.} simulations, respectively.  The images are centred on the most massive galaxy and each show a $1$ $\Mpc$ by $1$ $\Mpc$ (physical) region.  The \model{Smag.} model shows more substructure than the \model{None} and \model{Grad.} models.  The \model{Grad.} model shows many extended tails from substructure moving through the hot halo compared to the control and \model{Smag.} models, in addition to an overall smoother distribution in velocity space.}
    \label{fig:velocity_proj}
\end{figure*}

While the velocity power spectra results in Section~\ref{sec:driven_turbulence_spectra} show that eddy viscosity is required in Lagrangian finite mass methods, we find no significant impact of eddy viscosity on the \textit{average} gas and galaxy properties in our cosmological simulations. However, we should expect these results considering that the eddy viscosity models we tested in Section~\ref{sec:driven_turbulence_spectra} had no impact on the largest scales of the simulation (as they should not).  In terms of global gas properties, we investigated the vorticity, temperature, and density distributions of the warm-hot intergalactic medium (WHIM) and circumgalactic medium (CGM) phases\footnote{Definitions in Section~\ref{sec:cosmo_global}.} and found only minor differences in data binned by galaxy stellar mass.  Additionally, there were little differences between stellar mass distributions in our small volume, low resolution tests.  To compare the impact of eddy viscosity between the models we need to investigate the small scales of the simulations in a controlled manner.  For that reason, we will use the most massive galaxy in our cosmological simulations as a qualitative case study of the impact of the eddy viscosity model on the halo gas since this galaxy represents the same system in each simulation we ran.

First, we investigate the temperature projections of the most massive (in stellar mass) galaxy at redshifts $z = 2$, $1$, and $0$.  We confirmed that the most massive galaxy at $z = 2$ ends up as part of the most massive galaxy at $z = 1$ and $z = 0$.  We restrict our analysis to a radius of $500$ $\kpc$ (physical) for the purposes of this introductory study.  Additionally, given that we see the \model{Smag.}, \model{Grad.}, and \model{Dyn. Grad.} models as the best choices  from the results in Section~\ref{sec:cosmo_global}, we restrict our analysis to the \model{None}, \model{Smag.}, and \model{Grad.} models.  The \model{Grad.} model is much less computationally expensive than the \model{Dyn. Grad.} model and is, therefore, the better choice for cosmological simulations\footnote{Further study is required for tests that rely on getting the small scale structure as correct as possible, such as cold clouds interacting with a hot medium.}.

Fig.~\ref{fig:temperature_proj} shows the density-weighted temperature projections of the most massive galaxy at $z = 2$, $1$, and $0$ in rows from top to bottom, respectively.  The columns show the \model{None}, \model{Smag.}, and \model{Grad.} models from left to right, respectively. 

First, we compare the results at $z = 2$ across mixing model variants.  The dark clumps in the \model{None} case are individual cold galaxies that are being fed into the main structure via filaments.  Ongoing stellar and AGN feedback leads to the temperature increase in the central region, while the extended $T \sim 10^6$ $\mathrm{K}$ halo is a mixture of gravitational heated gas and gas heated by previous generations of stellar feedback and AGN.  There are many filamentary structures visible feeding the main galaxy (centred), and there is some small scale structure visible surrounding the galaxies.  The \model{Smag.} case  resembles the \model{None} case, 
except there less cold gas in the upper in-falling structure.  Additionally, the satellite galaxies have more small-scale structure in the cold gas surrounding them.  Importantly, the distribution of hotter gas appears mores smoothly distributed throughout the volume due to an increase in conversion of kinetic energy in thermal energy via the eddy viscosity.  

The temperature projection of the \model{Grad.} model simulation is strikingly different at $z = 2$ than both the \model{Smag.} and \model{None} cases.   While in the \model{None} and \model{Smag.} cases there appears to be large-scale gas at $T \lesssim 10^5$ $\mathrm{K}$ at the boundary of the $500$ $\kpc$ radius, there is no such gas in the \model{Grad.} case.  There is much less small-scale structure in the \model{Grad.} case, and the filaments are much smoothly distributed in space.  We conclude that the stellar feedback, gravitational heating, and AGN feedback are much more effective at heating the gas on small scales in the \model{Grad.} simulation, since the eddy viscosity is the strongest.

At $z = 1$, in the middle row of Fig.~\ref{fig:temperature_proj}, a similar picture emerges.  The \model{Smag.} case has much more small scale structure than the \model{None} case as is evident in the central region and to the leftmost satellite galaxy structure.  There is still a much smoother distribution of cold gas that extends outwards from the satellite galaxies.  There are only a few centrally-located cold gas clumps in the \model{Grad.} case compared with the \model{Smag.} case and there is effectively no fragmentation identifiable in the satellite galaxy structure to the left of panel.  

The last row of Fig.~\ref{fig:temperature_proj} shows $z = 0$ across the three model comparisons.  By this redshift, there is little structure remaining in the group-sized halo and the temperature distribution of the intragroup gas is very smooth.  Similarly to the other redshifts, the \model{Smag.} shows the most fragmentation of cold gas, followed by the \model{None} case, and then the \model{Grad.} case.  All three show a satellite being stripped of gas in the upper right of the panels but the \model{Smag.} and \model{Grad.} models each differ in an important unique way compared to the \model{None} case.   

It is much easier to see the differences in ram pressure stripping in velocity space rather than temperature space.  Fig.~\ref{fig:velocity_proj} shows the density-weighted velocity magnitude projections of the most massive galaxy at $z = 0$ for the three eddy viscosity models \model{None}, \model{Smag.}, and \model{Grad.} from left to right, respectively.  It is apparent across all three cases that the substructure is moving at least a factor of $\approx 5$ faster than the background gas yet only the \model{Grad.} case shows the most clearly defined long stripping tails from the cold gas in the galaxies.  The \model{Smag.} case does not have the cold gas in the satellite galaxy marked by the arrows in the \model{None} and \model{Grad.} cases, as the gas has been completely stripped away.  The \model{Grad.} case produces the cold gas in that satellite galaxy, and the tail is much more extended than in the \model{None} case.  In fact, it is evident upon close inspection of the cold gas structures in the halo that the \model{Grad.} case produces tails from the cold gas in the galaxies more readily than the \model{None} or \model{Smag.} cases.  That has implications for the study of ram pressure stripping in general, and should be further investigated in the future.

\section{Conclusions}
\label{sec:conclusions}

Turbulence is a key physical process in the study galaxy evolution and one of many highly complex non-linear interactions that must be understood to advance our knowledge of the Universe.  The complexity demands the use of simulations that combine astrophysical sub-grid models with hydrodynamics and gravity in an expanding universe.  All hydrodynamical simulations are known to require additional sub-grid models to accurately treat the impact sub-grid turbulence, yet these models have been widely ignored in the astrophysical community.  We have, for the first time in Lagrangian hydrodynamics, implemented and studied the gradient model \citep{Clark1979} -- an anisotropic sub-grid turbulence model for viscosity and metal mixing \citep{Hu2020}.  The model is based on directly modelling the error terms that arise from discretisation of the fluid field via a Taylor series expansion, including the compression terms that are missing from the standard Smagorinsky model.  We additionally implemented a dynamic procedure that computes the model parameter on-the-fly for the gradient model following the approach of \cite{Rennehan2019}.  We used the mesh-free finite mass method in the \pkg{GIZMO} code \citep{Hopkins2015a} as our numerical hydrodynamics solver for all of our experiments.

We ran driven turbulence simulations at Mach numbers $\mathcal{M} = 0.3$, $0.7$, and $2.1$ to validate the gradient model and compare with the popular Smagorinsky model.   \cite{Hu2020} recently showed, by post-processing driven their turbulence simulations, that the gradient model should be able to reduce the build-up of kinetic energy near the resolution scale in isotropic, homogeneous turbulence and better reproduce the sub-grid metal flux.  We confirmed these results in Section~\ref{sec:driven_turbulence} by using the gradient model at simulation time. 

Our analysis of the velocity power spectra in driven turbulence produced unexpected results.  We found that the gradient and Smagorinsky models, and their dynamic variants, predicted insufficient dissipation to reduce the artificial build-up of kinetic energy near the grid-scale in our $256^3$ simulations.  For that reason we introduced a boost factor $\gamma$ for the dissipation strength (i.e. $||\mvec{K}||\rightarrow \gamma ||\mvec{K}||$) and experimented with various values in the range $\gamma \in [1, 100]$.  We found that a factor of $\gamma \sim 10$ is sufficient to reduce the build-up of kinetic energy and is required for all of the Smagorinsky and gradient model variants.  Additionally, we found that the boost factor only needs to be applied to the subsonic ($\mathcal{M} < 1$) particles in our simulations to produce the correct statistics in supersonic turbulence (our $\mathcal{M} \sim 2.1$ test).  The true boost factor is higher since we used the maximum interaction distance between neighbouring gas particles to calculate the diffusion tensor, leading to a $\sim4$ times \textit{additional} boost (total $\sim40$) over other the default implementation in \pkg{GIZMO} \citep{Hopkins2018}.

Our converged metal mixing simulations in Section~\ref{sec:driven_turbulence_metals} show that when we use the gradient and Smagorinsky models at lower resolution ($64^3$ particles) we are able to produce MDFs that are equivalent to $4$ to $12$ times the resolution.  That is true for both the constant-coefficient and dynamic variants of the gradient and Smagorinsky models with standard parameter values, with the additional factor of $\sim4$ boost from using the maximum interaction distance in the kernel.  However, lower calibrations such as those from the \pkg{FIRE} simulations do not produce the correct rate of mixing as they are at least a factor of $\sim20$ too low.  We posit that this is because of the common calibration approach in cosmological simulations: calibration in tandem with the full suite of astrophysical sub-grid models.  

We argue that calibration of fundamental hydrodynamics models, such as the metal mixing model here, \emph{must be done in the absence of sub-grid astrophysical models}.   There must be a strong hydrodynamics base before the complexity of astrophysics is built on top.  Note that our dynamic Smagorinsky and dynamic gradient models \emph{do not require calibration} and produce accurate predictions for the rate of metal mixing in isotropic, homogeneous turbulence. 

As a first application of the new gradient model, we investigated a set of cosmological simulations to determine if there is any dominant impact on the galaxy and gas properties.  We investigated the metal mixing and eddy viscosity separately in Sections~\ref{sec:cosmo_global} \& ~\ref{sec:cosmo_viscosity}, respectively.

We found that the choice of sub-grid metal mixing model strongly impacts the MDF evolution in the warm-hot intergalactic medium (WHIM) and circumgalactic medium (CGM).  We found that the gradient and dynamic gradient models mix metals much more rapidly than the Smagorinsky variants and produce tighter MDFs up until $z \sim 1$ when they approach a similar distribution down to $z = 0$.  In our simulation without sub-grid metal mixing, the MDFs in the WHIM and CGM are significantly broader than any of the simulations with sub-grid metal mixing demonstrating that, at the very minimum, including any sub-grid metal mixing model is an improvement.  The most important result we discovered is that the metal mixing models are most impactful during the tempestuous early stages of galaxy evolution.  On very long timescales, the equilibrium distributions match quite closely across the models.

Including eddy viscosity in our cosmological simulations did not significantly impact the galaxy properties we investigated when averaging in bins of stellar mass after $z = 2$.  The galaxy stellar mass function was relatively unchanged, along with only slight variations in the stellar mass to halo mass function.  We also found that including eddy viscosity did not significantly impact the averaged gas distributions of vorticity, temperature, and density across galaxies of similar stellar mass.  We expected \textit{a priori} that the large-scale properties of galaxies would be unaffected as the sub-grid eddy viscosity mainly impacts the small-scale.   For that reason, we investigated a single system that could be linked across all of our cosmological simulations to gain a qualitative view of the impact.

In Section~\ref{sec:cosmo_viscosity} we showed the temperature projections of the most massive halo traced from $z = 2$, $1$ and $0$ in our cosmological simulations for three eddy viscosity models: no model, the standard Smagorinsky model, and the new constant-coefficient gradient model.  We found that the Smagorinsky model produced much more fragmentation in the halo gas of the most massive galaxy on the small scale compared to  having no eddy viscosity, at all redshifts.  We also found that the spatial temperature distribution was much smoother at $z = 2$ when stellar and active galactic nuclei feedback was much stronger, showing that the small-scale kinetic energy was being efficiently converted into thermal energy.  Although the constant-coefficient gradient model seemingly dissipates faster based on our results in Section~\ref{sec:driven_turbulence_spectra}, we observed that its inherent anisotropy does not lead to the same fragmentation we saw in the Smagorinsky model.  At high redshift, $z = 2$ the gradient model produced a much more widespread hot gas.  The filamentary structure at all redshifts was much smoother and, after $z = 1$, the satellite galaxies in the halo had many more clearly defined tails due to an improved treatment of ram-pressure stripping.  

Sub-grid metal mixing and eddy viscosity models have a strong impact on galaxy evolution simulations.  In this work, we showed in the simplest case of isotropic, homogeneous turbulence that the all of the models tested here improved the accuracy of metal mixing and turbulent kinetic energy dissipation in the mesh-free finite mass method.  The most significant differences between model choice appeared at high redshift in the early stages of galaxy evolution, before any equilibrium is reached.  Given that contemporary cosmological simulations have resolutions of less than $\sim50^3$ particles in a typical galaxy, we recommend that future studies must at least use the constant-coefficient gradient model as it is (a) computationally inexpensive compared to the dynamic version, while producing similar results and (b) includes the full velocity tensor in the diffusion tensor to give the most accurate solution for sub-grid turbulence.  Our recommendation is especially pertinent given the recent push to study higher redshift systems driven by the upcoming launch of the \textit{James Webb Space Telescope}, as our theoretical understanding of enrichment and thermodynamic histories will depend directly on sub-grid turbulence model choice.

\section*{Acknowledgements}

This research was enabled in part by support provided by WestGrid and Compute/Calcul Canada.  The simulations in this research were made possible by SciNet and the Niagara supercomputing cluster. DR acknowledges the support of the Natural Sciences and Engineering Research Council of Canada (NSERC), [funding reference number 534263] and through the Discovery Grant program.  DR thanks Arif Babul, Belaid Moa, Ondrea Clarkson, Austin Davis, Drummond Fielding, Phil Hopkins, and Wolfram Schmidt for useful advice during the course of this research.  DR gives special acknowledgement to Arif Babul and Belaid Moa for providing invaluable support to this research, without which it would not have been possible.  DR also thanks Daniel Price for his helpful comments during the reviewing stage that improved this work.

Our analysis was performed using the Python programming language (Python Software Foundation, \url{https://www.python.org}).  The following packages were used throughout the analysis: \pkg{h5py} \citep{Collette2013}, \pkg{numpy} \citep{Harris2020}, \pkg{scipy} \citep{Virtanen2020}, \pkg{yt} \citep{Turk2011}, and \pkg{matplotlib} \citep{Hunter2007}.  Prototyping of the analysis scripts was performed in the IPython environment \citep{Perez2007}.

\section*{Data availability}

The data underlying this article will be shared on reasonable request to the authour.




\bibliographystyle{mnras}
\bibliography{main} 



\appendix

\section{FILTERING APPROXIMATION}
\label{app:filterapprox}

Consider a scalar field $f_i(\mvec{r})$ where $\mvec{r}=(x,y,z)$ in a general coordinate system.  We expand $f_i(\mvec{r}')$ about $\mvec{r}$ in equation~(\ref{eq:filteroperation}) via a Taylor expansion,

\begin{equation}
\begin{split}
\label{eq:generictaylor}
f_i(\mvec{r}') = f_i(\mvec{r}) + [(\Delta\mvec{R})\cdot \mvec{\nabla}f(\mvec{r})] + \frac{1}{2}[(\Delta\mvec{R})\otimes (H(\mvec{r})\cdot (\Delta\mvec{R}))] \\ + \mathcal{O}(|\Delta\mvec{R}|^3),
\end{split}
\end{equation}

\noindent where $\Delta\mvec{R} = \mvec{r}'-\mvec{r}$ and $H(\mvec{r})$ is the Hessian matrix. Putting this into equation \ref{eq:filteroperation} gives,

\begin{equation}
\begin{split}
\label{eq:taylorintegrate}
\overline{f}_i(\mvec{r}) = f_i(\mvec{r}) + \frac{1}{2}H(\mvec{r})\int_{\mathrm{kern}} ((\Delta\mvec{R})\otimes (\Delta\mvec{R}))G(|\Delta\mvec{R}|, h) d\mvec{r}' \\ + \mathcal{O}(|\Delta\mvec{R}|^4).
\end{split}
\end{equation}

\noindent using the fact that the kernel function is normalized and the integral of an odd function over the domain is zero. The integral in equation \ref{eq:taylorintegrate} can be tabulated for a specific function $G=G(|\Delta\mvec{R}|, h)$, so let's define,

\begin{equation}
\label{eq:epsilonvector}
\epsilon(\mvec{r}) \equiv \frac{1}{2} \int_{\mathrm{kern}} ((\Delta\mvec{R})\otimes (\Delta\mvec{R}))G(|\Delta\mvec{R}|, h) d\mvec{r}'.
\end{equation}

\noindent We then have,

\begin{equation}
\label{eq:taylorvector}
\overline{f}_i(\mvec{r}) \approx f_i(\mvec{r}) + \epsilon(\mvec{r})\cdot H(\mvec{r}).
\end{equation}

\noindent However, $\epsilon(\mvec{r})$ is isotropic $\epsilon(\mvec{r}) = \epsilon\, \mvec{I}$, so we can write

\begin{equation}
\label{eq:secondmomentapprox}
\overline{f}_i(\mvec{r}) \approx f_i(\mvec{r}) + \epsilon\mvec{\nabla}^2 f_i(\mvec{r}).
\end{equation}


\bsp	
\label{lastpage}
\end{document}